%% file: main.tex
\documentclass[10pt,sigconf]{acmart}

\usepackage{graphicx}
\usepackage{balance}
\usepackage{url}
\usepackage{subfigure}
\usepackage{listings}
\usepackage{caption}
\usepackage[ruled]{algorithm}
\PassOptionsToPackage{noend}{algpseudocode}
\usepackage{algpseudocode}

\renewcommand{\algorithmiccomment}[1]{\bgroup\hfill//~#1\egroup}

\algnewcommand\algorithmicswitch{\textbf{switch}}
\algnewcommand\algorithmiccase{\textbf{case}}

\algdef{SE}[SWITCH]{Switch}{EndSwitch}[1]{\algorithmicswitch\ #1\ \algorithmicdo}{\algorithmicend\ \algorithmicswitch}%
\algdef{SE}[CASE]{Case}{EndCase}[1]{\algorithmiccase\ #1}{\algorithmicend\ \algorithmiccase}%
\algtext*{EndSwitch}%
\algtext*{EndCase}%

\newtheorem{theorem}{Theorem}
\newtheorem{proposition}{Proposition}

\setcopyright{none}
\renewcommand\footnotetextcopyrightpermission[1]{}
\begin{document}
\fancyhead{}

\title{Property Graph Schema Optimization for Domain-Specific Knowledge Graphs}

\author{Chuan Lei$^1$, Rana Alotaibi$^2$, Abdul Quamar$^1$, Vasilis Efthymiou$^1$, Fatma \"Ozcan$^1$}
\affiliation{%
  \institution{IBM Research - Almaden$^1$, University of California at San Diego$^2$}
}
\email{chuan.lei|vasilis.efthymiou@ibm.com,ahquamar|fozcan@us.ibm.com,ralotaib@eng.ucsd.edu}

\begin{abstract}
Enterprises are creating domain-specific knowledge graphs by curating and integrating their business data from multiple sources. The data in these knowledge graphs can be described using ontologies, which provide a semantic abstraction to define the content in terms of the entities and the relationships of the domain. The rich semantic relationships in an ontology contain a variety of opportunities to reduce edge traversals and consequently improve the graph query performance. Although there has been a lot of effort to build systems that enable efficient querying over knowledge graphs, the problem of schema optimization for query performance has been largely ignored in the graph setting. In this work, we show that graph schema design has significant impact on query performance, and then propose optimization algorithms that exploit the opportunities from the domain ontology to generate efficient property graph schemas. To the best of our knowledge, we are the first to present an ontology-driven approach for property graph schema optimization. We conduct empirical evaluations with two real-world knowledge graphs from medical and financial domains. The results show that the schemas produced by the optimization algorithms achieve up to 2 orders of magnitude speed-up compared to the baseline approach. 
\end{abstract}

\maketitle

\input{intro}
\input{background}
\input{rules}
\input{baseline}
\input{optimizer}
\input{exp}

\input{related_work}
\input{conclusions}

\balance

\bibliographystyle{abbrv}
\bibliography{main.bib}

\appendix
\input{appendix1}
\input{appendix2}

\end{document}

%% file: intro.tex
\section{Introduction}
\label{sec:intro}

Domain-specific knowledge graphs are playing an increasingly important role to derive business insights in many enterprise applications such as customer engagement, fraud detection, network management, etc~\cite{Noy:2019}. One distinct characteristic of these enterprise knowledge graphs, compared to the open-domain knowledge graphs like DBpedia~\cite{LehmannIJJKMHMK15}, Freebase~\cite{Bollacker08freebase}, and YAGO2~\cite{Suchanek:2008}, is their deep domain specialization. The domain specialization is typically captured by an ontology which provides a semantic abstraction to describe the entities and their relationships of the data in the knowledge graphs. A few widely used domain-specific ontologies include Unified Medical Language System (UMLS)\footnote{\url{https://www.nlm.nih.gov/research/umls/index.html}} and SNOMED Clinical Terms\footnote{\url{http://www.snomed.org/}} in the medical domain, Financial Industry Business Ontology (FIBO)\footnote{\url{https://spec.edmcouncil.org/fibo/}} and Financial Report Ontology (FRO)\footnote{\url{http://www.xbrlsite.com/2015/fro/us-gaap/xbrl/Ontology/Overview.html}} in the financial domain, and many more in various other domains\footnote{\url{https://lod-cloud.net/}}.
The ontology is often used to drive the creation of a knowledge graph by ingesting and transforming raw data from multiple sources into standard terminologies. The curated knowledge graphs allow users to express their queries in standard vocabularies, which promotes more interoperable and effective enterprise applications and services for specific domains~\cite{DBLP:series/synthesis/2015Dong,DBLP:series/synthesis/2015Christophides}.

There are two popular approaches to store and query knowledge graphs: RDF data model and SPARQL query language~\cite{DBLP:journals/sigmod/SakrA09} or property graph model and graph query languages such as Gremlin~\cite{DBLP:conf/sigmod/SunFSKHX15} and Cypher~\cite{DBLP:conf/sigmod/FrancisGGLLMPRS18}. An important difference between RDF and property graphs is that RDF regularizes the graph representation as a set of triples, which means that even literals are represented as graph vertices. Such artificial vertices make it hard to express graph queries in a natural way. The property graph model instead uses vertices to represent entities and edges to represent the relationships between them, with each specified using key-value properties pairs~\cite{DBLP:conf/grades/RestHKMC16}. For this reason, property graph systems are rapidly gaining popularity for graph storage and retrieval. Examples include Neo4j~\cite{neo4j}, Apache JanusGraph~\cite{janus}, Azure Cosmos DB~\cite{cosmos}, Amazon Neptune~\cite{neptune}, to name a few. Many graph applications (e.g., community detection, centrality analysis, and link prediction) heavily rely on the performance of graph queries over the property graph systems. Many techniques have been proposed for optimizing the query performance, system scalability, and transaction support for these systems~\cite{DBLP:conf/icde/NeumannM11,DBLP:conf/icde/MeimarisPMA17,DBLP:conf/edbt/TsialiamanisSFCB12,DBLP:conf/sigmod/BorneaDKSDUB13}. However the problem of property graph schema optimization has been largely ignored, which is also critical to graph query performance.

In this paper, we tackle the \textit{property graph schema optimization problem} for domain-specific knowledge graphs. Our goal is to create an optimized schema\footnote{We use the terms property graph schema, graph schema, and schema interchangeably.} based on a given ontology, such that the corresponding property graph can efficiently support various types of graph queries (e.g., pattern matching, path finding, or aggregation queries) with better query performance. The raw data is loaded directly as a property graph that conforms to the optimized schema\footnote{A property graph schema may not be logically equivalent to a given ontology. Capturing the full expressivity of ontologies (e.g., negation, role inclusion, transitivity) in the form of a property graph schema is an unexplored and challenging problem, which is beyond the scope of this work.}. One straightforward way to create a property graph schema from an ontology is to directly map each ontology concept to a schema node, and to map each ontology relationship to a schema edge, analogous to ER diagram to relational schema mapping. However, we argue that the graph query performance varies vastly for different property graphs with the same data but corresponding to different schemas, and the rich semantic information in the ontology provides unique opportunities for schema optimization. We illustrate this using two examples from the medical domain.

\begin{figure}[!htb]
\centering
\subfigure[Snippet of a Medical Ontology]{
\includegraphics[width=0.9\columnwidth]{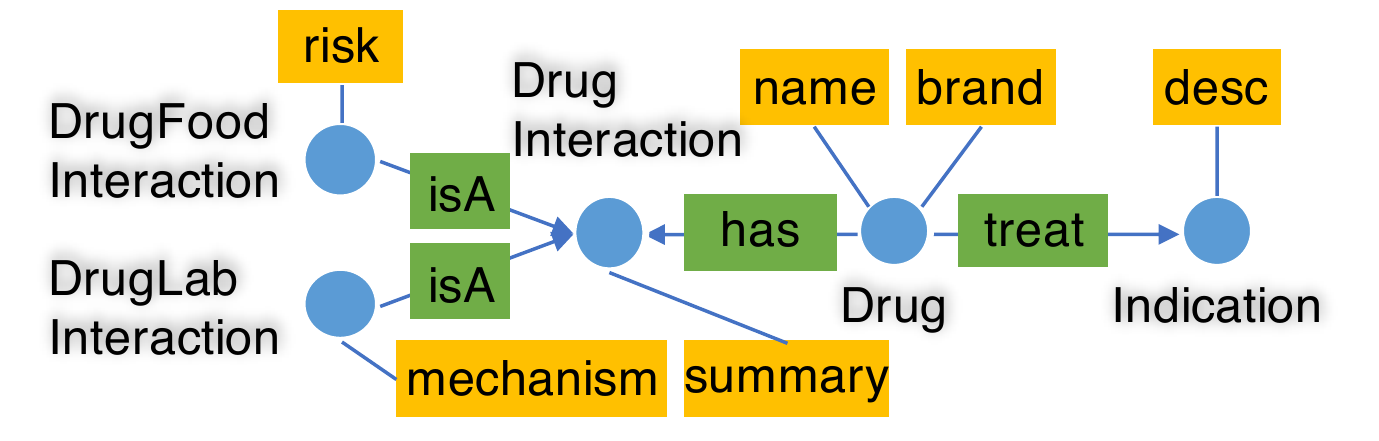}  
\label{fig:eg1_ontology}
}\vspace{-5pt}
\subfigure[Property Graph 1 (Direct)]{
\includegraphics[width=0.9\columnwidth]{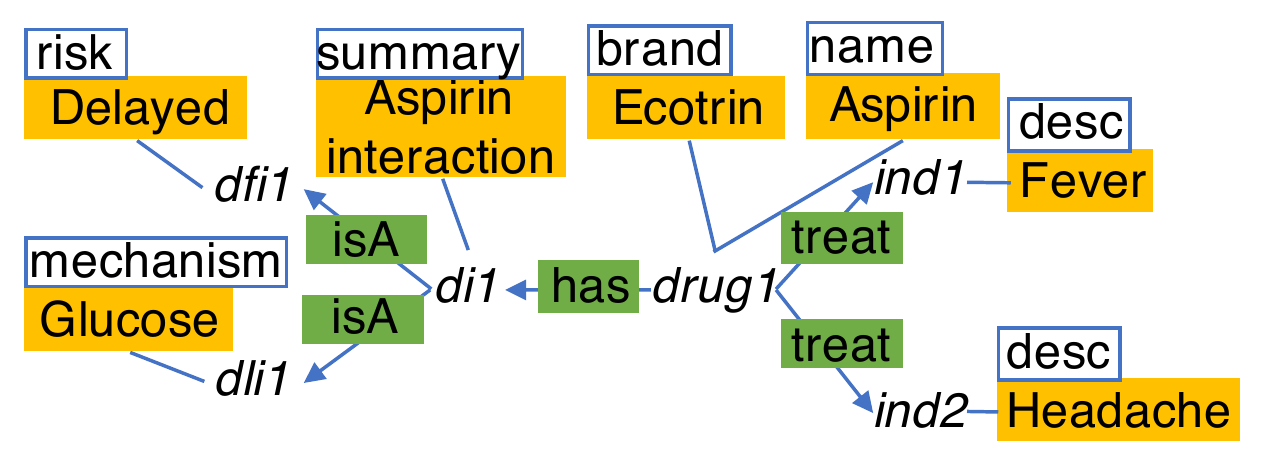}
\label{fig:eg1_direct}
}\vspace{-5pt}
\subfigure[Property Graph 2 (Optimized)]{
\includegraphics[width=0.9\columnwidth]{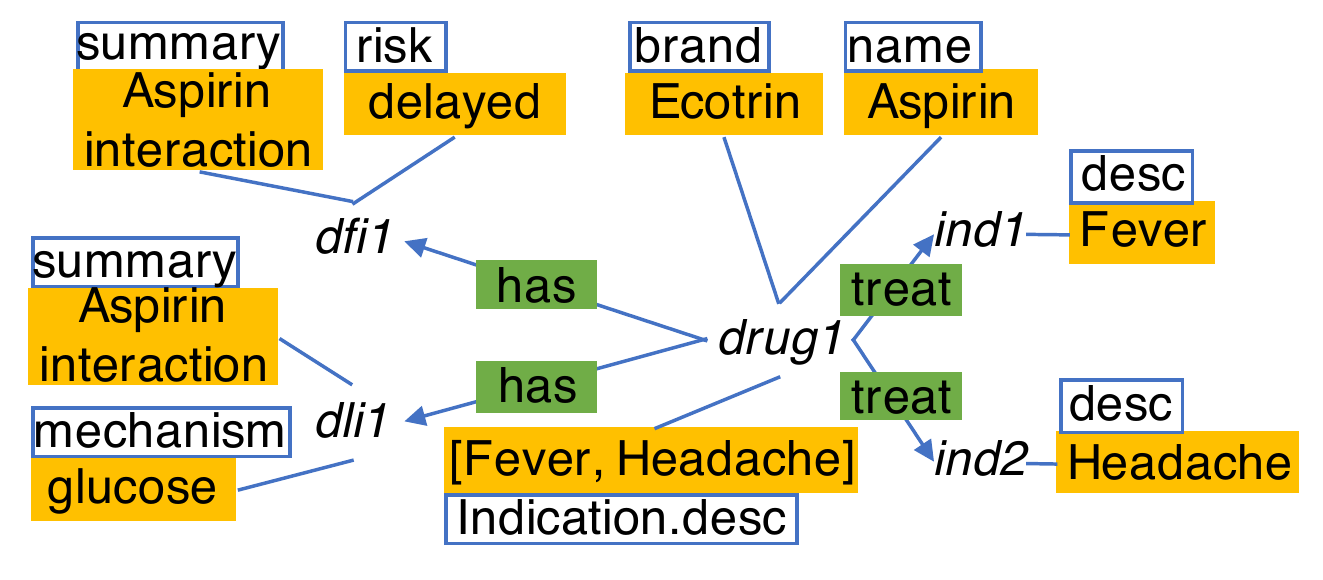}
\label{fig:eg1_opt}
}
\caption{Motivating Example.} 
\label{fig:eg1}
\end{figure}

\textit{Example 1 (Pattern matching query).} Consider the ontology in Figure~\ref{fig:eg1_ontology}, {\it summary} is a property of {\it DrugInteraction} concept, which is connected to {\it DrugFoodInteraction} and {\it DrugLabInteraction} concepts via inheritance ({\it isA}) relationships. Figures~\ref{fig:eg1_direct} and~\ref{fig:eg1_opt} show two alternative property graphs conforming to two different schemas with several vertices and edges. In Figure~\ref{fig:eg1_direct}, the vertex \textit{di1} (i.e., an instance of \textit{DrugInteraction}) leads to both \textit{dfi1} and \textit{dli1}. In Figure~\ref{fig:eg1_opt}, \textit{drug1} directly connects to {\it dfi1} and {\it dli1} vertices. For any query that requires edge traversals from \textit{drug1} to either \textit{dfi1} or \textit{dli1} or both, the property graph 2 clearly requires less number of edge traversals. A pattern matching query interested in \textit{Drug} and the associated \textit{risk} of \textit{DrugFoodInteraction} achieves 2 orders of magnitude performance gains on the optimized property graph (23ms) compared to the property graph 1 (3245ms).

\textit{Example 2 (Aggregation query).} In Figure~\ref{fig:eg1_ontology}, {\it Drug} concept is also connected to {\it Indication} concept via a \textit{treat} ($1$:$M$) relationship. In this case, we observe that if we replicate certain properties accessible via a \textit{1:M} relationship, edge traversals can be avoided. Figure~\ref{fig:eg1_opt} shows that the vertex \textit{drug1} has an additional property, which is a list of descriptions replicated from the property {\it desc} of {\it ind1} and {\it ind2}. An aggregation query (COUNT) on the \textit{desc} of {\it Indication} treated by {\it Drug} runs 8 times faster on this optimized property graph (78ms) than the property graph 1 (627ms). In this case, avoiding the edge traversals is extremely beneficial, especially when the number of edges between these two types of vertices is large.

These two examples show that edge traversal is one of the dominant factors affecting graph query performance, and having an optimized schema can greatly improve query performance. We can reduce edge traversals by merging nodes or replicating data. However, this needs to be done carefully, as the resulting knowledge graph needs to preserve its semantics information. Fortunately, the rich semantic relationships in an ontology provide a variety of opportunities to reduce graph traversals. To generate an optimized graph schema, we need to identify and exploit these opportunities in the ontology, and design different techniques to utilize them accordingly. As illustrated in the examples, certain optimization techniques require data replication resulting in space overheads. Hence, the schema optimization has to trade off between the query performance and the space consumption of the resulting property graph.

{\bf Our proposed approach.} To the best of our knowledge, we are the first to address the problem of property graph schema optimization to improve graph query performance. In addition to the ontology, our approach also takes into account the space constraints, if any, and additional information such as data distribution and workload summaries\footnote{We refer to the access frequency of concepts, relationships and properties as workload summaries which will be formally defined later.}. We propose a set of rules that are designed to optimize the graph query performance with respect to different types of relationships in the ontology. When there is a space constraint, we estimate the cost-benefit of applying these rules to each individual relationship by leveraging the additional data distribution and workload information. We propose two algorithms, concept-centric and relation-centric, which incorporate the cost-benefit scores to produce an optimized property graph schema. 
Our approach can seamlessly handle updates to the property graph, as long as its schema remains unchanged.

{\bf Contributions.} The contributions of this paper can be summarized as follows:

1. We introduce an ontology-driven approach for property graph schema optimization.
	
2. We design a set of rules that reduce the edge traversals by exploiting the rich semantic relationships in the ontology, resulting in better graph query performance.
	
3. We propose concept-centric and relation-centric algorithms that harness the proposed rules to generate an optimized property graph schema from an ontology, under space constraints. The concept-centric algorithm utilizes the centrality analysis of concepts, and the relation-centric algorithm uses a cost-benefit model.
	
4. Our experiments show that our ontology-driven approach effectively produces optimized graph schemas for two real-world knowledge graphs from medical and financial domains. The queries over the optimized property graphs achieve up to 2 orders of magnitude performance gains compared to the graphs resulting from the baseline approach.

The rest of the paper is organized as follows. Section~\ref{sec:background} introduces the basic concepts, formulates the problem, and provides an overview of our ontology-driven approach. Section~\ref{sec:rules} describes our optimization rules for different types of relationships in an ontology. Section~\ref{sec:baseline} explains the algorithms to produce optimized property graph schema. We provide our experimental results in Section~\ref{sec:exp}, review related work in Section~\ref{sec:related}, and finally conclude in Section~\ref{sec:conclusion}.

%% file: background.tex
\section{Preliminaries \& Approach Overview}
\label{sec:background}

\subsection{Preliminaries}
\label{sec:background:defs}

An ontology describes a particular domain and provides a structured view of the data. Specifically, it provides an expressive data model for the concepts that are relevant to that domain, the properties associated with the concepts, and the relationships between concepts.

\begin{definition}[Ontology ($\mathcal{O}$)]
An ontology $\mathcal{O}$ ($C$, $R$, $P$) contains a set of concepts $C =\{c_n | 1 \leq n \leq N\}$, a set of data properties $P = \{p_m | 1\leq m \leq M\}$, and a set of relationships between the concepts $R=\{r_k|1 \leq k \leq K\}$.
\label{def:ontology_graph}
\end{definition}

An ontology is typically described in OWL~\cite{owl}, wherein a concept is defined as a {\em class}, a property associated with a concept is defined as a {\em DataProperty} and a relationship between a pair of concepts is defined as an {\em ObjectProperty}. Each DataProperty $p_i \in P_n$ represents a characteristic of a concept $c_n \in C$ and $P_n \subseteq P$ represents the set of DataProperties associated with the concept $c_n$. Each ObjectProperty $r_k=(c_s,c_d,t)$ is associated with a source concept $c_s \in C$, also referred to as the domain of the ObjectProperty, a destination concept $c_d \in C$, also referred to as the range of the ObjectProperty, and a type $t$. The type $t$ can be either a functional (i.e., {\it 1:1}, {\it 1:M}, {\it M:N}), an inheritance (a.k.a {\it isA}) or a union/membership relationship\footnote{Even if inheritance and union are not ObjectProperties, we simplify the notation for presentation purposes.}. In this paper, we use the ontology as a semantic data model of a knowledge graph.

\begin{figure}[!htb]
\centering
\includegraphics[width=0.9\columnwidth]{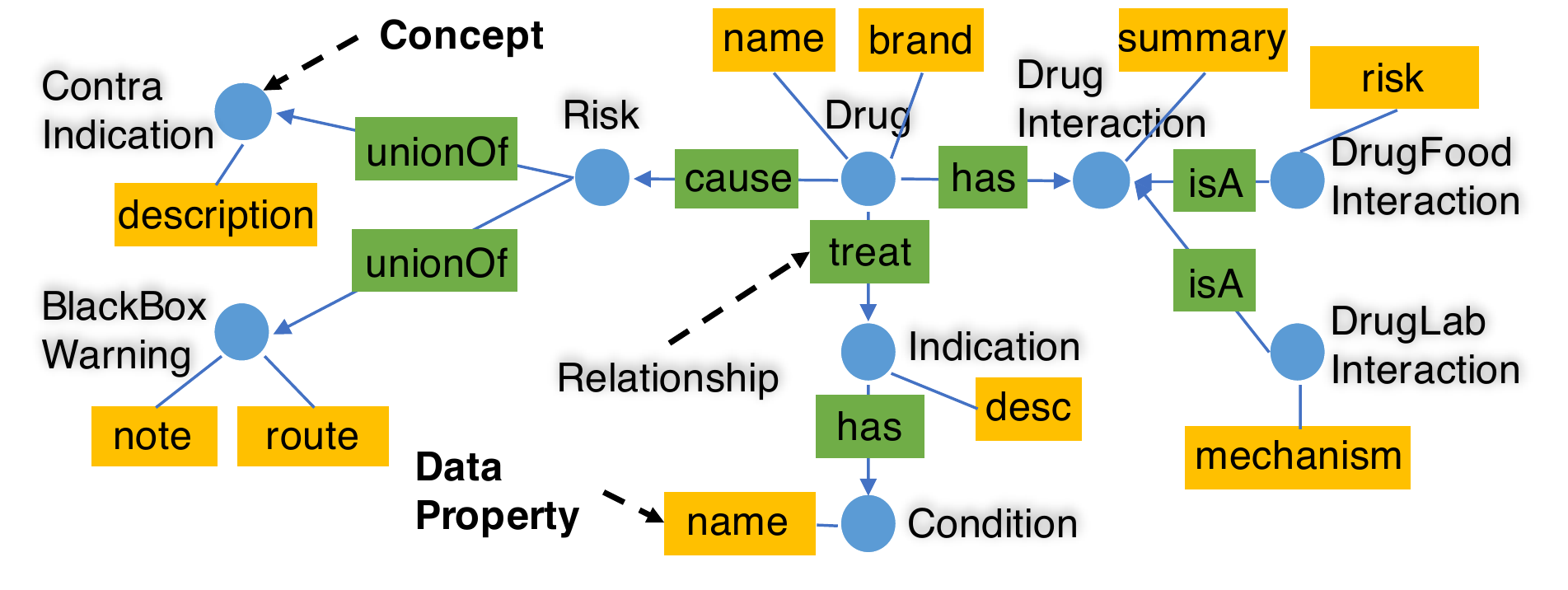}  
\caption{Medical Ontology.}
\label{fig:ontology}
\end{figure}

We adopt the widely used property graph model from~\cite{Robinson:2013}.

\begin{definition}[Property Graph ($\mathcal{PG}$)]
A property graph $\mathcal{PG}$ ({\it V, E}) is a directed multi-graph with vertex set $V$ and edge set $E$, where each node $v\in V$ and each edge $e\in E$ has data properties consisting of multiple attribute-value pairs.
\label{def:property_graph_schema}
\end{definition}

Similar to a relational database schema that describes tables, columns, and relationships of a relational database, the property graph schema is critical for creating high-quality domain-specific graphs. A property graph instantiated from a property graph schema provides agile and robust knowledge services with correctness, coverage, and freshness~\cite{Noy:2019}.

A property graph schema $\mathcal{PGS}$ can be specified in a data definition language such as Neo4j's Cypher~\cite{DBLP:conf/sigmod/FrancisGGLLMPRS18}, TigerGraph's GSQL~\cite{DBLP:journals/corr/abs-1901-08248}, or GraphQL SDL~\cite{Hartig:2019:DSP:3327964.3328495}. They all define notions of node types and edge types, as well as property types that are associated with a node type or with an edge type. We adopt Cypher due to its popularity, but our proposed techniques are independent of the aforementioned languages. Table 1 provides the notations used in this paper.

\begin{table}[!htb]
\caption{Notations.}
\centering
\begin{tabular}{|l||l|}
\hline
Notations & Definitions \\
\hline\hline
$\mathcal{O}$ & an ontology \\
\hline
$c_i$ & $c_i\in C$: a concept in an ontology \\
\hline
$r_i$ & $r_i\in R$: a relationship in an ontology \\
\hline
$c_i.P_i$ & all data properties associated to $c_i$ \\
\hline
$c_i.inE$ & all incoming relationships of $c_i$ \\
\hline
$c_i.outE$ & all outgoing relationships of $c_i$ \\
\hline
$c_i.R_i$ & $c_i.R_i$ = $c_i.inE$ $\cup$ $c_i.outE$ \\
\hline
$r_i.src$ & the source concept of $r_i$ \\
\hline
$r_i.dst$ & the destination concept of $r_i$ \\
\hline
$r_i.type$ & the relationship type of $r_i$ (i.e., {\it 1:1}, $union$, \\ 
           & $inheritance$, {\it 1:M}, or {\it M:N}) \\
\hline\hline
$\mathcal{PGS}$ & a property graph schema \\
\hline
$vs_i$ & $vs_i\in VS$: a schema vertex \\
\hline
$vs_i.PS_i$ & all property schema of $vs_i$ \\
\hline
$es_i$ & an edge schema defined in $\mathcal{PGS}$ \\
\hline
$es_i.type$ & the edge type of $e_i$ \\
\hline\hline
$\mathcal{PG}$ & a property graph \\
\hline
$V_i$ & $V_i\in V$ all instance vertices of $vs_i$ \\
\hline
$v_{i,j}$ & $v_{i,j}\in V_i$, an instance vertex of $vs_i$\\
\hline
$v_{i,j}.p_k$ & a property of $v_{i,j}$\\
\hline
$e_k$ & $e_k = (v_{src}, v_{dst}) \in E$, $v_{src}$, $v_{dst}\in V$ \\
\hline
\end{tabular}
\label{tab:notations}
\end{table}

\subsection{Approach Overview}
\label{sec:background:overview}

Given an ontology $\mathcal{O}$ providing a semantic abstraction of the input data, the problem of \textit{property graph schema optimization} is to generate a property graph schema that produces the best query performance for various graph queries (e.g., pattern matching, path finding, or aggregation queries). Optimizing the property graph might entail data replication and hence increased memory footprint. In real knowledge graph applications, especially in a multi-tenant setting,  there is a limit on the amount of memory that we can trade for query performance. Hence, any practical solution needs incorporate a space constraint while producing an optimized property graph schema.

\begin{figure}[!htb]
\centering
\includegraphics[width=0.9\columnwidth]{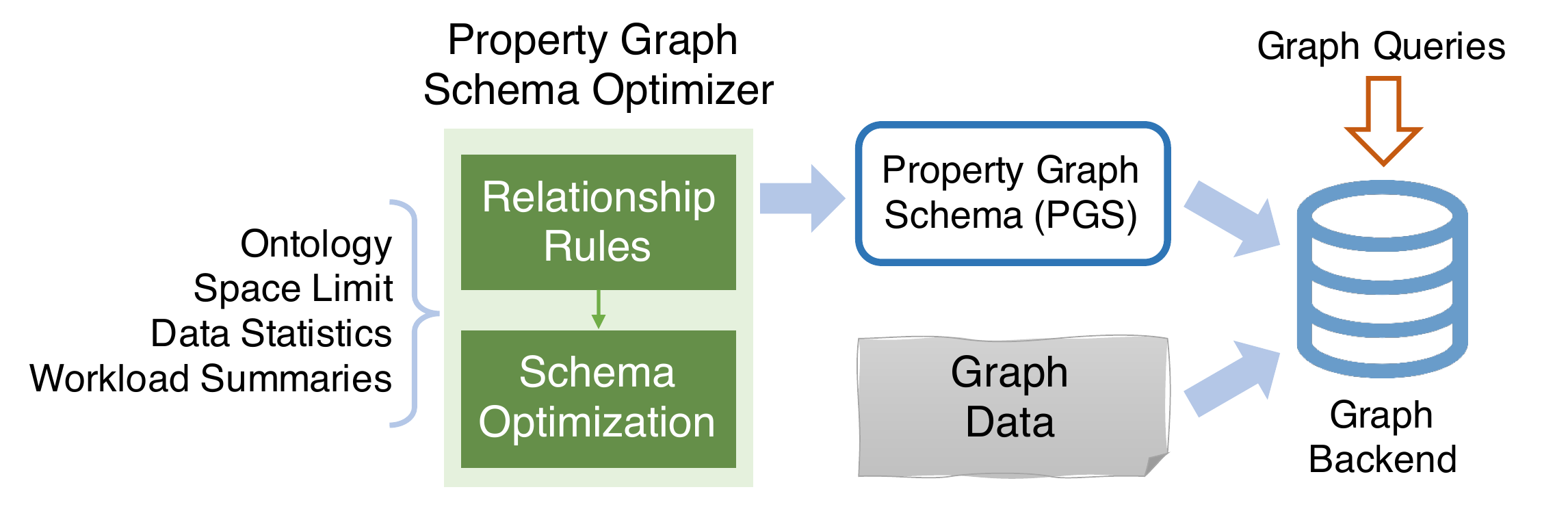}  
\caption{Approach Overview.}
\label{fig:system}
\end{figure}

Figure~\ref{fig:system} provides an overview of our property graph schema optimization approach. The property graph schema optimizer takes as input an ontology and optionally a space limit, data statistics, as well as workload summaries\footnote{Access frequencies of concepts, relationships, and data properties in an ontology}. It utilizes a set of rules designed for different types of relationships to produce an optimized property graph schema. The raw graph data is then loaded into a graph database (e.g., Neo4j or JanusGraph) conforming to the optimized schema. At query time, users can directly expresses graph queries against this instantiated property graph corresponding to the optimized schema.

%% file: rules.tex
\section{Relationship Rules}
\label{sec:rules}

Graph queries often involve multi-hop traversal or vertex attribute lookup/analytics on property graphs. As shown in the motivating examples, edge traversals over a graph are vital to the overall query performance. Hence, we focus on the rich semantic relationships in an ontology and propose a set of novel rules for different types of relationships. These rules minimize edge traversals and consequently improve graph query performance.

{\bf Union Rule.} In an ontology, a union relationship ($r_{un} = (c_i, c_j)$) contains a union concept ($c_i$) and a member concept ($c_j$). Each instance of a union concept is an instance of one of its member concepts, and each instance of a member concept is also an instance of the union concept. Figure~\ref{fig:ontology} shows that \textit{BlackBoxWarning} and \textit{ContraIndication} are two member concepts of a union concept \textit{Risk}. A graph query accessing an instance of \textit{Risk} is equivalent to accessing the instances of either \textit{BlackBoxWarning}, or \textit{ContraIndication}, or both. In other words, if we create a property graph directly from the ontology shown in Figure~\ref{fig:ontology}, then the queries starting from any vertices of either \textit{BlackBoxWarning} or \textit{ContraIndication} concepts have to traverse through some vertex of \textit{Risk} in order to reach the vertices of \textit{Drug}. This leads to unnecessary edge traversal. 

\begin{figure}[!htb]
\centering
\subfigure[Optimized PGS]{
\includegraphics[width=0.55\columnwidth]{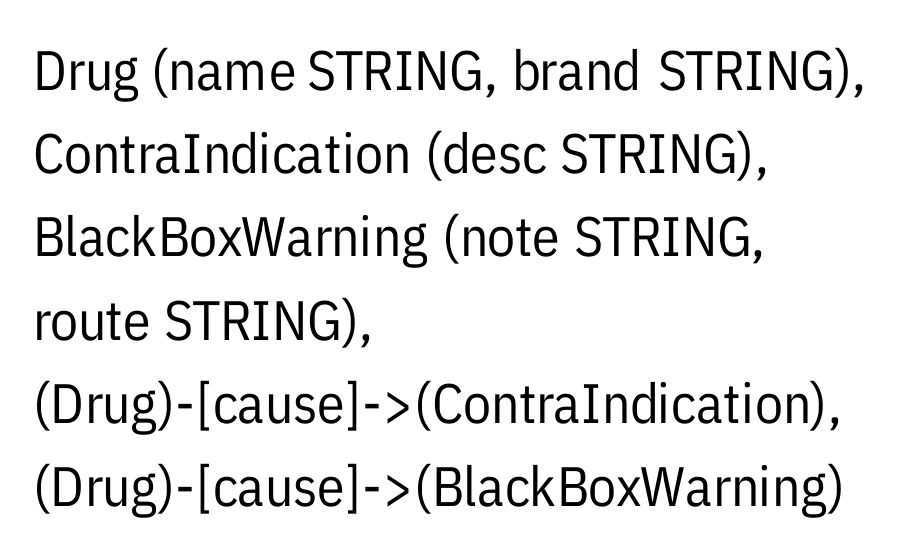}  
\label{fig:union1}
}
\subfigure[Optimized PG]{
\includegraphics[width=0.34\columnwidth]{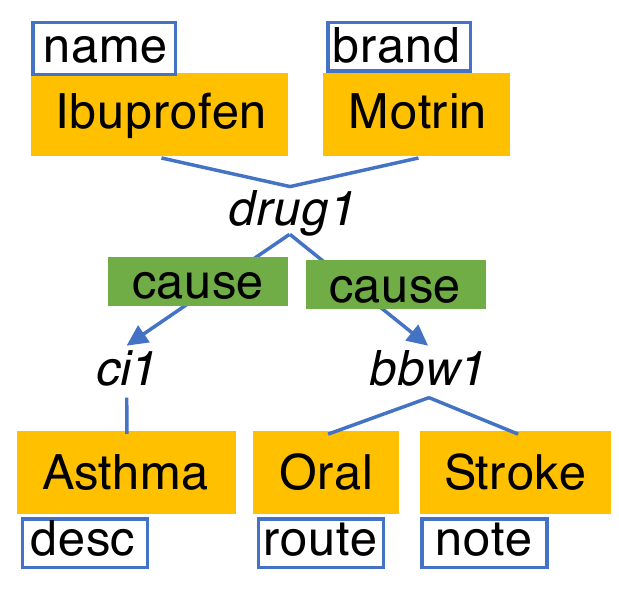}
\label{fig:union2}
}
\caption{Union Relationship.} 
\label{fig:union}
\end{figure}

\begin{algorithm}[!htb]
\caption{Union Rule (union)}
\label{algo:union:nospace}
\begin{algorithmic}[1]
\Require A union relationships $r_{un}$ 
\State $vs_{i} \leftarrow r_{un}.src$
\Comment{the union concept of $r_{un}$}
\State $vs_{j} \leftarrow r_{un}.dst$ 
\Comment{the member concept of $r_{un}$}
\ForAll {$r \in vs_{i}.ES_i$}
    \If {$\neg (r$ of type $union$)} 
	    \State $vs_{j}.ES_j \leftarrow vs_{j}.ES_j \cup r$
    \EndIf
\EndFor
\end{algorithmic}
\end{algorithm}

Hence we propose a union rule to alleviate this issue. The union rule first creates a union node $vs_i$ (based on the corresponding $c_i$ in $\mathcal{O}$) and its member node $vs_j$ (based on the corresponding $c_j$ in $\mathcal{O}$) in the property graph schema. Then the member node $vs_j$ is connected to the other nodes that connect to the union node $vs_i$ in the property graph schema (Algorithm~\ref{algo:union:nospace}). Figures~\ref{fig:union1} and~\ref{fig:union2} show the property graph schema and the corresponding property graph after applying the union rule to the above example. In the optimized property graph, retrieving the drugs (e.g., \textit{Ibuprofen}) that cause \textit{Asthma} requires only a single edge traversal, instead of 2 in the property graph directly instantiated from the ontology.

{\bf Inheritance Rule.} An inheritance relationship ($r_{ih} = (c_i, c_j)$) contains a parent concept ($c_i$) and a child concept ($c_j$). Similar to the union rule, we create a parent node $vs_i$ (corresponding to $c_i$) and its child node $vs_j$ (corresponding to $c_j$) in the property graph schema. Unlike a union concept, a parent concept in the inheritance relationship may have instances that are not present in any of its children concepts.

\begin{enumerate}
    \item Connect the child node $vs_j$ directly to the nodes that are connected to its parent node $vs_i$, and attach all data properties $vs_i.P_i$ of $vs_i$ to the child node $vs_j$ in the schema;
    
    \item Connect the parent node directly to the nodes that are connected to its child node, and attach all data properties $vs_j.P_j$ of $vs_j$ to the parent node $vs_i$ in the schema;
    
    \item Or connect the parent $vs_i$ and child $vs_j$ nodes with an edge of type \textit{isA}.
\end{enumerate}

\begin{figure}[!htb]
\centering
\subfigure[Optimized PGS 1]{
\includegraphics[width=0.5\columnwidth]{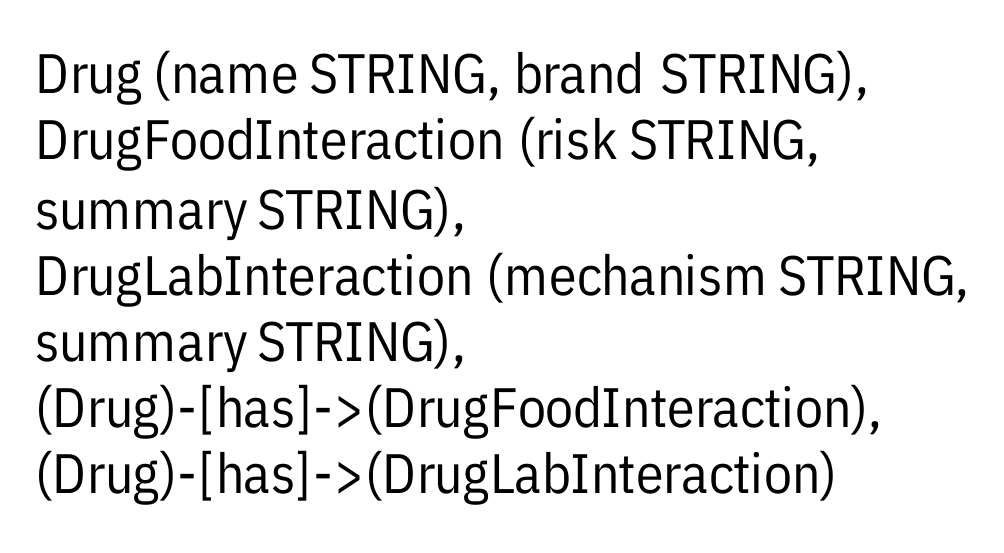}  
\label{fig:isa1}
}\hspace{-10pt}
\subfigure[Optimized PG 1]{
\includegraphics[width=0.44\columnwidth]{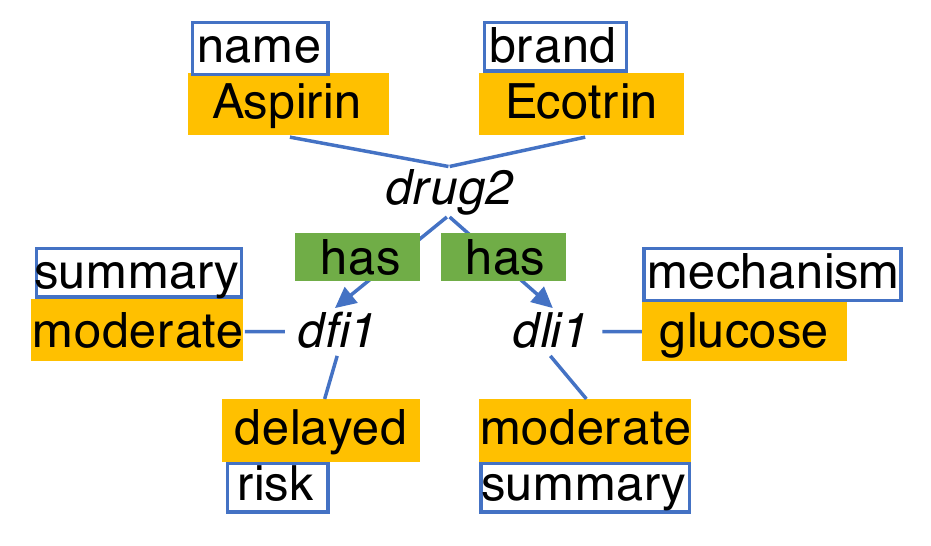}
\label{fig:isa2}
}
\subfigure[Optimized PGS 2]{
\includegraphics[width=0.5\columnwidth]{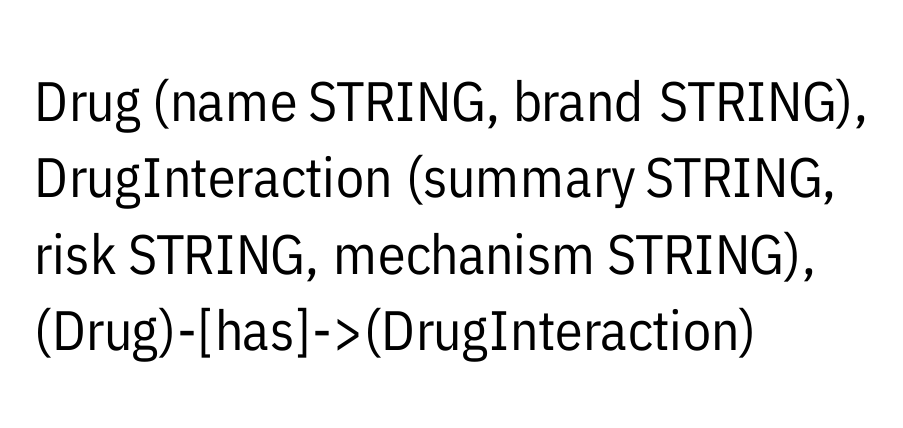}
\label{fig:isa3}
}\hspace{-10pt}
\subfigure[Optimized PG 2]{
\includegraphics[width=0.44\columnwidth]{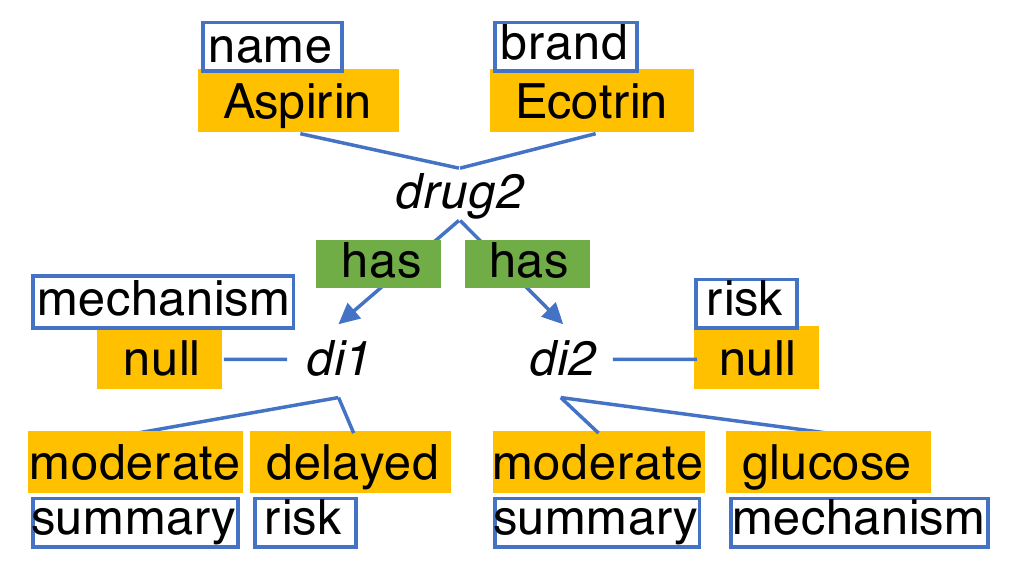}
\label{fig:isa4}
}
\caption{Inheritance Relationship.} 
\label{fig:isa}
\end{figure}

In the first two cases, edge traversals can be avoided in the property graph conforming to the property graph schema. Figure~\ref{fig:ontology} shows that \textit{DrugFoodInteraction} and \textit{DrugLabInteraction} are two children concepts of \textit{DrugInteraction}. Applying the inheritance rule to these concepts can lead to two alternative optimized property graph schemas shown in Figure~\ref{fig:isa}. Figures~\ref{fig:isa1} and~\ref{fig:isa2} demonstrate the first scenario where the data properties (\textit{summary}) of the parent concept \textit{DrugInteraction} are directly attached to two children concepts \textit{DrugFoodInteraction} and \textit{DrugLabInteraction}.  Figures~\ref{fig:isa3} and~\ref{fig:isa4} depict the second scenario where the data properties \textit{risk} and \textit{mechanism} of two respective children concepts are now attached to the parent concept \textit{DrugInteraction}.

However, attaching the data properties ($c_i.P_i$) from the parent concept to the child concept incurs data replication as $c_i.P_i$ is shared among all children concepts (Figure~\ref{fig:isa2}). If the number of data properties shared by the children concepts is large, the data replication can introduce significant space overhead. On the other hand, when the data properties ($c_j.P_j$) from the children concepts are replicated to their parent concept ($c_i$), $c_i$ may end up with a large number of data properties (Figure~\ref{fig:isa4}). However, these data properties may not exist in many instance vertices of $c_i$. Consequently, the instance vertices of $c_i$ may consume unnecessary space. To remedy the above two issues, we propose to exploit the Jaccard similarity~\cite{Leskovec:2014} between $c_i.P_i$ and $c_j.P_j$ to decide the best strategy for the inheritance relationship:

\begin{equation}
\label{eq:jaccard}
JS(c_i.P_i,\ c_j.P_j) = |c_i.P_i \cap c_j.P_j|\ /\ |c_i.P_i \cup c_j.P_j|. 
\end{equation}

As described in Algorithm~\ref{algo:isa:nospace}, if $JS(c_i.P_i, c_j.P_j)$ is greater than a threshold $\theta_1$, it indicates that the child concept $c_j$ shares a lot of data properties with its parent concept $c_i$. Intuitively, this means that $c_j$ has only few properties in addition to the ones of $c_i$. 
In this case, moving $c_j.P_j$ from the child concept to $c_i$ incurs less space overhead compared to the other way. Similarly, if $JS(c_i.P_i, c_j.P_j)$ is less than a threshold $\theta_2$ ($\theta_2$ $\leq$ $\theta_1$), the child concept $c_j$ has little in common with its parent $c_i$. Intuitively, this means that $c_j$ has many additional properties compared to $c_i$. Therefore, it is more cost effective to make the data properties of the parent concept $c_i.P_i$ available at $c_j$. In either case, the inheritance rule avoids edge traversals in the resulting property graph. 

\begin{algorithm}[!htb]
\caption{Inheritance Rule (inheritance)}
\label{algo:isa:nospace}
\begin{algorithmic}[1]
\Require An inheritance relationship $r_{ih}$
\State $vs_{i} \leftarrow r_{ih}.src$\Comment{Parent concept}
\State $vs_{j} \leftarrow r_{ih}.dst$\Comment{Child concept}
\State $jsim \leftarrow JS(vs_i.PS_i,vs_j.PS_j)$
\Comment{Jaccard similarity of $r_{ih}$}
\If {$jsim > \theta_1$} 
    \State $vs_{i}.P_i \leftarrow vs_{i}.PS_i \cup vs_{j}.PS_j$
    \newline\Comment{\hspace*{2mm}$ES_{ih}$ is the set of inheritance relationships}
    \State $vs_{i}.ES_i \leftarrow (vs_{i}.ES_i \cup vs_{j}.ES_j$)\textbackslash $r_{ih}$
\ElsIf {$jsim < \theta_2$}
    \State $vs_{j}.PS_j \leftarrow vs_{j}.PS_j \cup vs_{i}.PS_i$
    \State $vs_{j}.ES_j \leftarrow (vs_{j}.ES_j \cup vs_i.ES_i$)\textbackslash $r_{ih}$
\EndIf
\end{algorithmic}
\end{algorithm}

Note that the similarity score of a parent concept and a child concept remains unchanged even if new data properties are added to one or both concepts as a result of applying other rules. The reason is that the Jaccard similarity is computed based on the given ontology, as it represents the semantic similarity between two concepts with an inheritance relationship. Hence we calculate the Jaccard similarity score for all inheritance relationships before applying any rules.

\begin{figure}[!htb]
\centering
\subfigure[Optimized PGS]{
\includegraphics[width=0.48\columnwidth]{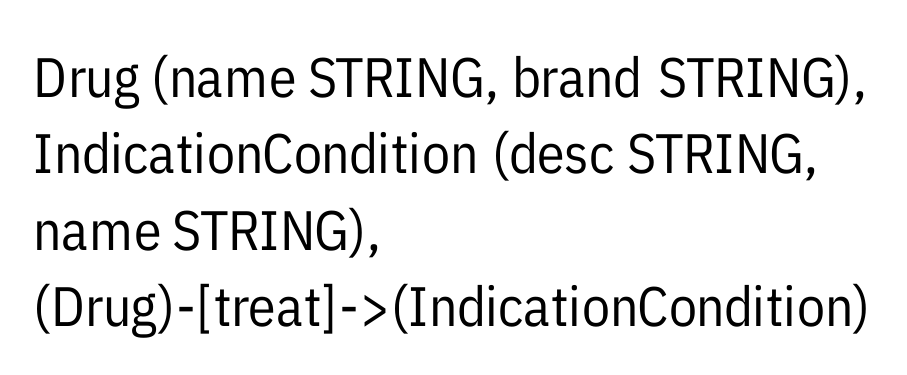}  
\label{fig:oneone1}
}\hspace{-5pt}
\subfigure[Optimized PG]{
\includegraphics[width=0.42\columnwidth]{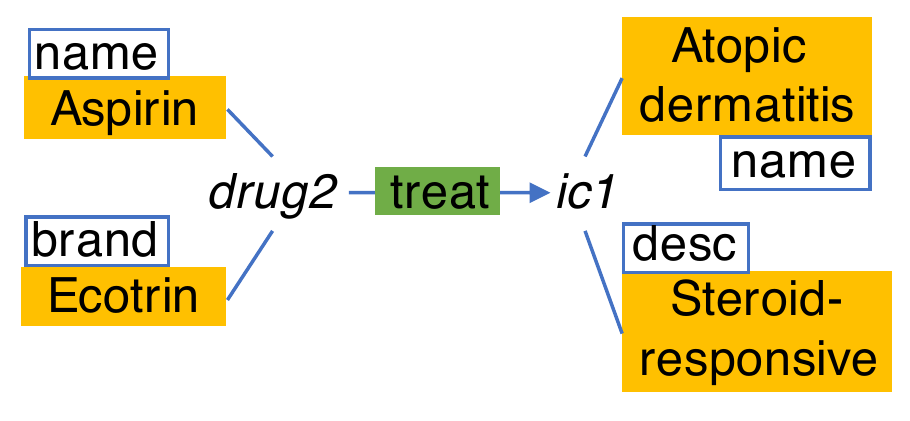}
\label{fig:oneone2}
}
\caption{1:1 Relationship.} 
\label{fig:oneone}
\end{figure}

{\bf One-to-one Rule.} A {\it 1:1} relationship ($r_{1:1} = (c_i, c_j)$) indicates that an instance of $c_i$ can only relate to one instance of $c_j$ and vice versa (e.g., \textit{Indication} and \textit{Condition} in Figure~\ref{fig:ontology}). Two concepts ($c_i$ and $c_j$) of a {\it 1:1} relationship can be represented as one combined node $vs_{ij}$ in the optimized schema, which is similar to joining two tables in relational databases where one row in one table is linked with only one row in another table and vice versa. If two tables are merged, then a join can be saved when two tables are queried together. The other tables can still join with the merged table through their respective relationships. Namely, the {\it 1:1} rule preserves the original semantics and does not lead to any information loss. Any query accessing instance vertices of $c_i$ and $c_j$ can be satisfied by looking up the merged instance vertex of $c_{ij}$. In Figure~\ref{fig:oneone1}, \textit{IndicationCondition} is the merged concept with two data properties, \textit{name} and \textit{note}, attached. Hence the edge traversal (e.g., from \textit{Drug} to \textit{Condition} in Figure~\ref{fig:ontology}) is avoided and the number of instance vertices (i.e., space consumption) is reduced as well. Algorithm~\ref{algo:1to1:nospace} shows the one-to-one rule, which is straightforward to follow.

\begin{algorithm}[!htb]
\caption{1:1 Rule (oneToOne)}
\label{algo:1to1:nospace}
\begin{algorithmic}[1]
\Require A $1$:$1$ relationship $r_{1:1}$ 
\State $vs_i \leftarrow r_{1:1}.src$
\State $vs_j \leftarrow r_{1:1}.dst$
\State $vs_{i,j} \leftarrow \emptyset$
\State $vs_{i,j}.ES_{i,j} \leftarrow (vs_i.ES_i \cup vs_j.ES_j$)\textbackslash $r_{1:1}$
\State $vs_{i,j}.PS_{i,j} \leftarrow vs_i.PS_i \cup vs_j.PS_j$
\end{algorithmic}
\end{algorithm}

{\bf One-to-many Rule.} A {\it 1:M} relationship ($r_{1:M}$ = ($c_i$, $c_j$)) indicates that an instance of $c_i$ can potentially refer to several instances of $c_j$). In other words, in a {\it 1:M} relationship, an instance of $c_i$ allows zero, one, or many corresponding instances of $c_j$. However, an instance of $c_j$ cannot have more than one corresponding instance of $c_i$.

\begin{figure}[!htb]
\centering
\subfigure[Optimized PGS]{
\includegraphics[width=0.5\columnwidth]{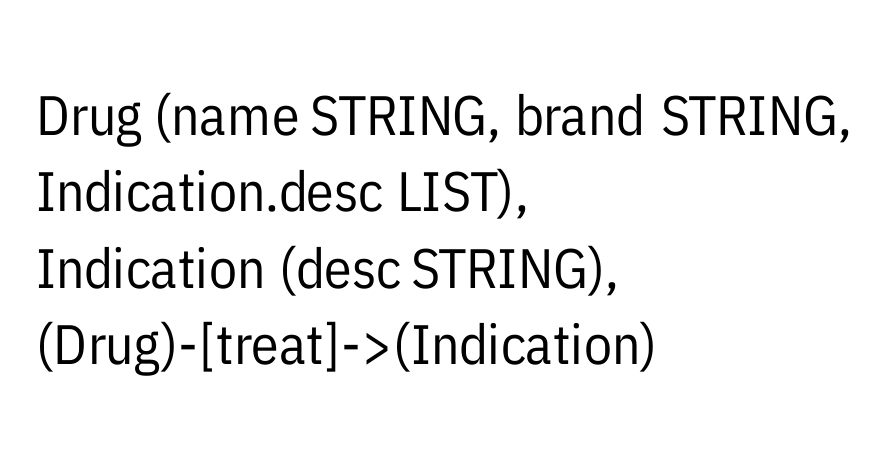}  
\label{fig:onemany1}
}\hspace{-5pt}
\subfigure[Optimized PG]{
\includegraphics[width=0.42\columnwidth]{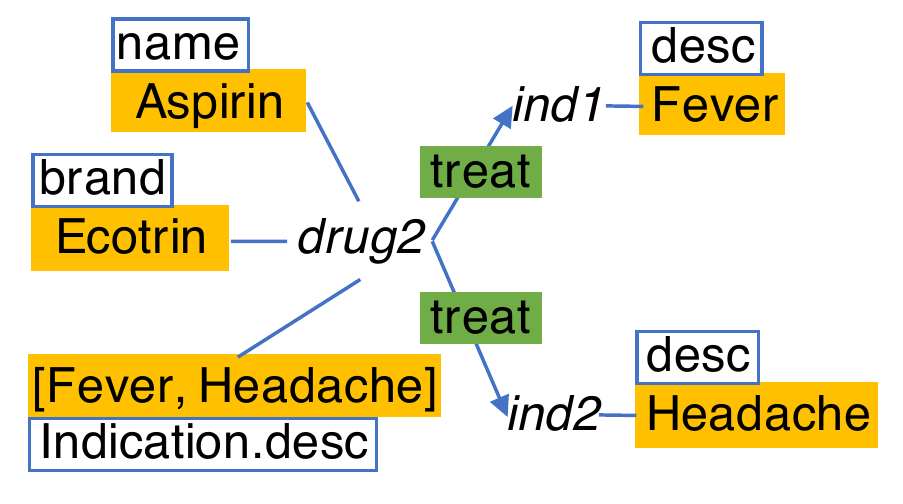}
\label{fig:onemany2}
}
\caption{1:M Relationship.} 
\label{fig:onemany}
\end{figure}

To better support the aggregation (e.g., COUNT, SUM, AVG, etc.) and neighborhood (1-hop) lookup functions in graph queries, we first create two nodes $vs_i$ and $vs_j$ corresponding to $c_i$ and $c_j$ in the optimized schema. Then we propagate each data property $vs_j.P_j$ of $vs_j$ as a property of type \textit{LIST} to the other node $vs_i$ (Fig.~\ref{fig:onemany1}). The aggregation and neighborhood lookup functions can directly leverage these localized list properties instead of traversing through the edges of the {\it 1:M} relationships. This is similar to denormalization technique in relational databases where data replication is added to one or more tables in order to avoid costly joins. As depicted in Figure~\ref{fig:onemany2}, \textit{Indication.desc} is a data property of \textit{drug2} consisting of a list of descriptions (i.e., \textit{[Fever, Headache]}) that saves the aggregation queries edge traversals to the other instance vertices (e.g., \textit{ind1} and \textit{ind2}). The potential savings can be substantial when there are many edges between instance vertices of two concepts such as \textit{Drug} and \textit{Indication}.

However, the newly introduced property of type $LIST$ introduces additional space overheads, which can be expensive depending on the data distribution. Therefore, choosing the appropriate set of data properties from each {\it 1:M} relationship to propagate is critical with respect to both query performance and space consumption. We will describe algorithms to choose the data properties to merge in Section~\ref{sec:opt}. Algorithm~\ref{algo:1toM:nospace} corresponds to the one-to-many rule.

\begin{algorithm}[h]
\caption{1:M Rule (oneToMany)}
\label{algo:1toM:nospace}
\begin{algorithmic}[1]
\Require A $1$:$M$ relationship $r_{1:M}$
\State $vs_i \leftarrow r_{1:M}.src$
\State $vs_j \leftarrow r_{1:M}.dst$
\ForAll {$p \in vs_j.PS_j$}
\State $vs_i.PS_i.$addAsList($p$)
\EndFor
\end{algorithmic}
\end{algorithm}

{\bf Many-to-many Rule.} An $M$:$N$ relationship ($r_{M:N}$ = ($c_{i}$, $c_{j}$)) indicates that an instance of $c_{i}$ can have several corresponding instances of $c_{j}$, and vice versa. An $M$:$N$ relationship is essentially equivalent to two $1$:$M$ relationships, namely, $r_{1:M} = (c_{i}, c_{j})$ and $r_{1:M} = (c_{j}, c_{i})$. Therefore, the many-to-many rule is identical to the one-to-many rule, except that the property propagation is done for both directions. Namely, in the optimized schema, a data property of the node $vs_i$ corresponding to $c_i$ in $\mathcal{O}$ is propagated as a property of type \textit{LIST} to the node $vs_j$ corresponding to $c_j$ in $\mathcal{O}$, and vice versa. Hence applying the many-to-many rule leads to the same potential gains for queries with aggregate or neighborhood (1-hop) lookup functions at the cost of introducing additional space consumption.

In summary, all proposed rules reduce the number of edge traversals which improve graph query performance. Moreover, these rules can be utilized in graph systems using different storage backends. The potential benefits could be more significant when the storage backend changes from in-memory to disk as edge traversals may incur additional disk I/Os. However, \textit{union}, \textit{inheritance}, \textit{one-to-many}, and \textit{many-to-many} rules may incur space overheads. In Section~\ref{sec:baseline}, we describe our property graph schema optimization algorithms, trading off performance gain and space overhead.

%% file: baseline.tex
\section{Property Graph Schema Optimization}
\label{sec:baseline}

In this section, we first introduce a property graph schema optimization algorithm in an ideal scenario (i.e., no space constraints). Then, we describe our concept-centric and relation-centric algorithms that harness the proposed rules and a cost-benefit model to generate an optimized property graph schema for a given space constraint.

\subsection{Optimization Without Space Constraints}

To produce an optimized property graph schema, we need to determine how to utilize the proposed rules described in Section~\ref{sec:rules}. A straightforward approach is to iteratively apply these rules in order and generate the property graph schema.

Specifically, Algorithm~\ref{algo:o2p:nospace} takes as input an ontology $O$ and first computes the Jaccard similarity scores for all inheritance relationships (Lines 1-2). Then, it iteratively applies the appropriate rule to each relationship in the ontology (Lines 3-16). At the end of each iteration, it checks if the ontology converges (Line 17). Finally when no more rule applies, a property graph schema is generated (Lines 18-19). In fact, these rules can be applied in any order, and the generated property graph schema is always the same.

\begin{algorithm}[!htb]
\caption{Ontology to PGS without Space Limits}
\label{algo:o2p:nospace}
\begin{algorithmic}[1]
\Require Ontology $\mathcal{O} = (C, R, P)$ 
\Ensure A property graph schema $\mathcal{PGS}\newline$
\hspace{90pt}\Comment{Compute Jaccard similarity for each inheritance relationship and get all parent concepts}
\ForAll {$r \in R$ of type $inheritance$}
    \State $r.js \leftarrow$ computeJS($r$)
\EndFor
\Repeat 
    \State $O_{prev} \leftarrow O$
    \ForAll{$r \in R$}
        \Switch{$r.type$}
            \Case{{\it 1:1}}
                \State $O \leftarrow$ oneToOne($O$, $r$)
            \EndCase
            \Case{{\it 1:M}}
                \State $O \leftarrow$ oneToMany($O$, $r$)
            \EndCase
            \Case{{\it M:N}}
                \State $O \leftarrow$ manyToMany($O$, $r$)
            \EndCase
            \Case{{\it union}}
                \State $O \leftarrow$ union($O$, $r$)
            \EndCase
            \Case{{\it inheritance}}
                \State $O \leftarrow$ inheritance($O$, $r$)
            \EndCase
        \EndSwitch
    \EndFor
\Until{$O = O_{prev}$}
\State $PGS \leftarrow$ generatePGS($O$)
\State \Return $PGS$
\end{algorithmic}
\end{algorithm}

\begin{theorem}
Applying the union, inheritance, {\it 1:M} and {\it M:N} rules in any order produces a unique 
$\mathcal{PGS}$, if there is no space constraint.
\label{theo:order}
\end{theorem}

The detailed proof can be found in Appendix~\ref{sec:appendix:2}.

%% file: optimizer.tex
\subsection{Schema Optimization With Space Constraints}
\label{sec:opt}

While the na\"ive approach harnesses all potential optimization opportunities aggressively, it incurs space overheads from {\it union}, {\it inheritance}, {\it 1:M}, and {\it M:N} rules. In cases where the number of such relationships is large in the ontology, this can be expensive with respect to the space consumption, especially in a cluster setting, where many large-scale property graphs co-exist. Hence our goal is to produce an optimized property graph schema for a given space limit. The quality and the space consumption of an optimized property graph schema are measured based on the total benefit and cost (i.e., space consumed) by applying the rules (given by Equations~\ref{eq:bc:union}-\ref{eq:bc:1m} in Section \ref{sec:opt:rc}).

\begin{definition}[Optimal Property Graph Schema]
Let $\mathbb{PGS}$ be the set of all property graph schemas, such that $\forall{\mathcal{PGS'}}\in\mathbb{PGS}$ we have $Cost(\mathcal{PGS'})\leq S$, where $S$ is a given space budget. $\mathcal{PGS}_{opt}\in\mathbb{PGS}$ is an optimal property graph schema if $\nexists$ $\mathcal{PGS'}$ $\in$ $\mathbb{PGS}$ such that Benefit($\mathcal{PGS'}$) $>$ $Benefit(\mathcal{PGS}_{opt})$.
\end{definition}

Finding an optimal property graph schema is exponential in the number of concepts and relationships in the ontology, which is practically infeasible. Hence, we need to design efficient heuristics to produce a near-optimal property graph schema. To achieve this goal, we propose two property graph schema optimization algorithms that leverage additional information such as data and workload characteristics.

{\bf Data characteristics} contain the basic statistics about each concept, data property, and relationship specified in the given ontology. The statistics include the cardinality of data instances of each concept and relationship, as well as the data type of each data property. The data characteristics allow us to identify and prioritize the more beneficial relationships when applying {\em union, inheritance, one-to-many} and {\em many-to-many} rules, such that the space can be used more efficiently.

{\bf Access frequencies} provide an abstraction of the workload in terms of how each concept, relationship, and data property accessed by each query in the workload. We use {\it AF($c_i\xrightarrow{r_k} c_j$.$P_j$)} to indicate the frequency of queries (the number of queries) that access a data property in $c_j.P_j$ from the concept $c_i$ through the relationship $r_k$. The high frequency of a relationship indicates its relative importance among all relationships in the given ontology. Hence it is more imperative to apply the above proposed rules to these relationships with high frequency.  

In case of no prior knowledge about access frequency, we assume that it follows a uniform distribution. Our approach can also handle updates (i.e., insert, delete, and modify) to the property graph if they do not incur any schema changes. If the accumulated updates change the data distributions, then we can apply the rules locally to the affected part of the ontology. Note that data statistics changes can invalidate certain rule applied earlier, or can trigger new rules, especially inheritance and union rules. We can make local adjustments to accommodate these changes. Minimizing such transformation overheads is left as future work.

\subsubsection{Concept-Centric Algorithm}
\label{sec:opt:cc}

As described in Section~\ref{sec:background}, an ontology describes a particular domain and provides a concept-centric view over domain-specific data. Intuitively, some concepts are more critical to the domain, and have more relationships with the other concepts~\cite{Abdul:2018}. We expect these key concepts to be queried more frequently than other concepts, which is confirmed in~\cite{DBLP:conf/sigmod/QuamarLMOKME20}. This leads to our concept-centric algorithm that exploits the structural information in an ontology to identify key concepts which we believe are more likely to be accessed more often. Hence, this algorithm is useful when no workload summary is available.

To determine these key concepts, we utilize centrality analysis over the ontology to rank all concepts according to their respective centrality score. The centrality analysis is based on the commonly used PageRank algorithm~\cite{Brin:1998} as its underlying assumption, more important websites likely to receive more links from other websites, is similar to our intuition of key concepts. Our modified PageRank algorithm, called $OntologyPR$ (Algorithm~\ref{algo:pagerank}), determines the centrality score of each concept in an ontology. Compared to PageRank, we further introduce weights for both in and out degrees of concepts in determining their centrality scores.

\begin{algorithm}[!htb]
\caption{Ontology PageRank Algorithm \\(OntologyPR)}
\label{algo:pagerank}
\begin{algorithmic}[1]
\Require $\mathcal{O} = (C, R, P)$
\Ensure $\mathcal{O} = (C, R, P)$
\State $C_{un} \leftarrow \text{empty set}$
\ForAll {$r \in R$ of type $union$} 
    \State $c_i \leftarrow r.src$
    \Comment{the union concept of $r$}
    \State $c_{j} \leftarrow r.dst$ 
    \Comment{the member concept of $r$}
    \State $C_{un}.$add$(c_i)$
    \State $c_j.R_j \leftarrow (c_j.R_j \cup c_i.R_i)$\textbackslash $r$
\EndFor
\State $O.$remove$(C_{un})$
\ForAll {$r \in R$}
    \If {$r$ is of type $inheritance$}
        \State $R_{ih}$.add$(r)$
        \State $O$.remove$(r)$
    \Else
        \State $O$.add$(r')$
        \Comment{add a reverse relation $r'$}
    \EndIf
\EndFor
\State pageRank($O$)
\Comment{PageRank on the modified ontology}
\State $O$.add$(R_{ih})$
\Comment{add inheritance relationships back}
\State updatePR($O$)
\Comment{update PageRank score for inheritance concepts}
\State \Return $O$
\Comment{$O$ associated with PageRank scores}
\end{algorithmic}
\end{algorithm} 

{\bf Inheritance.} To cater for inheritance relationships, we remove these relationships from the ontology while running the initial PageRank algorithm. This allows us to calculate the page ranks of a concept based on the links from other concepts that are not children of the same concept. After computing the page rank values of all concepts, we re-attach these relationships and update the page ranks of each concept by doing a depth-first traversal over its inheritance relationships to find the parent with the highest page rank. If this value is higher than the current page rank of the concept, we use this value as the new page rank of the concept. This enables a child concept to inherit the page rank of its parent. The intuition is that a child concept inherits all its other properties from the same chain of concepts and hence would have a similar estimate of centrality.

{\bf Unions.} The union concept in the ontology represents a logical membership of two or more concepts. Any incoming edge to a union concept can therefore be considered as pointing to at least one of the member concepts of the union. Similarly each outgoing edge can be considered as emanating from at least one of the member concepts.

To handle union concepts, the $OntologyPR$ algorithm iterates over all incoming and outgoing edges to/from the union concept. For each incoming edge to the union concept, we create new edges between the source concept and each of the member concepts of the union. For each outgoing edge, similarly, we create new edges between the destination and each of the member concepts of the union. Thus the page rank mass is appropriately distributed to/from the member nodes of the union. Finally, the union node itself is removed from the graph as its contribution towards centrality analysis has already been accounted for by the new edges to/from the member concepts of the union.

{\bf Out-degree of Concepts.} In the default PageRank algorithm, the weight distribution of the page rank is proportional to the in-degree of a node as it receives page rank values from all its neighbors that point to it. In other words nodes with a high in-degree would tend to have a higher page rank than nodes with a low in-degree. However, for a domain ontology, we observe that both in-degree and out-degree are equally important in terms of the key concept. Hence, we introduce a reverse edge in the ontology, essentially making the graph equivalent to an undirected graph. Then, the $OntologyPR$ algorithm uses this modified ontology as an input to determine the centrality score of each concept.

Using $OntologyPR$, we associate PageRank scores with each concept in the ontology. To accurately capture the relative importance of the concepts, we further leverage the {\em data characteristics} and {\em access frequency} information to rank all concepts. Namely, the ranking score for a concept is defined as follows.
\begin{equation}
\label{eq:conceptScore}
Score(c_i) = \frac{c_i.pr \times AF(c_i)}{Size(c_i)}
\end{equation}
\noindent where $c_i.pr$ denotes the PageRank score of $c_i$, $AF(c_i)$ denotes the access frequency of $c_i$ including accessing all data properties of $c_i$, and $Size(c_i)$ denotes the size of $c_i$ including all data properties of $c_i$.

Based on Equation~\ref{eq:conceptScore}, our concept-centric algorithm (Algorithm~\ref{algo:o2p:cc}) first sorts all concepts in a descending order of their respective scores (Lines 1-2). Then, it iterates through each concept $c$ (Lines 3-8). For each concept, the algorithm utilizes the $applyRules$ procedure to apply all rules (Section~\ref{sec:rules}) to the relationships connecting to $c$. During this process, the algorithm updates the space limit as it is consumed by the rules. Once the space is fully exhausted, the algorithm terminates (Lines 7-8) and returns the optimized property graph schema (Line 10).

\begin{algorithm}[!htb]
\caption{Concept-Centric Algorithm}
\label{algo:o2p:cc}
\begin{algorithmic}[1]
\Require Ontology $\mathcal{O} = (C, R, P)$, space limit $S$
\Ensure A property graph schema $\mathcal{PGS}$
\State $O\leftarrow$ ontologyPR($O$)
\State $C_{srt} \leftarrow$ sort($C$)
\ForAll {$c \in C_{srt}$}
    \ForAll {$r \in c.R$}
        \State $S' \leftarrow S$
        \State $O$, $S$ $\leftarrow$ applyRules($r$, $S'$)
        \If {$S < 0$} 
            \State \textbf{break} 
        \EndIf
    \EndFor
\EndFor
\State $PGS \leftarrow$ generatePGS($O$)
\State \Return $PGS$
\end{algorithmic}
\end{algorithm} 

{\bf Complexity Analysis.}
The \textit{OntologyPR} is the dominant procedure in Algorithm~\ref{algo:o2p:cc}, and its time complexity is $O((|R|+|C|)k)$, where $|R|$ is the number of relationships, $|C|$ is the number of concepts, and $k$ is the maximum number of iterations. The time complexity of sorting concepts is $O(|C|log|C|)$. Finally, the time complexity of applying rules to the sorted concepts is $O(|R|)$. Thus, the overall time complexity of Algorithm~\ref{algo:o2p:cc} is $O((|R|+|C|)k)$.

\subsubsection{Relation-Centric Algorithm}
\label{sec:opt:rc}

Intuitively, the concept-centric algorithm prioritizes the relationships of the key concepts in an ontology by leveraging information such as access frequency, data characteristics, and structural information from the ontology. However, the relationship selection is limited to each concept locally. Namely, the concept-centric algorithm does not have a global optimal ordering among all relationships in the ontology. To address this issue, we propose the relation-centric algorithm based on a cost-benefit model for each type of relationships described as follows.

{\bf Cost Benefit Models.} The union rule, introduced in Section~\ref{sec:rules}, connects the member concept directly to all concepts that are connected to the union concept. Then, the benefit of applying this rule to a union relationship $r$ is the access frequency of $r$, and the cost is the number of edges that we copy from the union concept to the member concept. Formally:
\begin{equation}
\label{eq:bc:union}
\begin{array}{ll}
Benefit(r) = AF(c_i\xrightarrow{r}c_j) \\
Cost(r) = \sum_{r'\in (c_i.R_i \backslash R_{un})}|r'|,
\end{array}
\end{equation}
\noindent where $c_i$ denotes the union concept and $|r'|$ denotes the number of edges between the instance vertices of $c_i$ and the ones of a neighborhood concept\footnote{The neighborhood concepts do not include the member concepts of $c_i$.} of $c_i$.

The benefit of applying the inheritance rule to an inheritance relationship is the access frequency of that relationship multiplied by the Jaccard similarity between $c_i.P_i$ and $c_j.P_j$. Depending on that similarity, the cost of inheritance rule can be either the number of new edges attached to the parent, or the number of new edges attached to the child. Formally: 
\begin{equation}
\label{eq:bc:ih}
\begin{array}{ll}
\hspace{-10pt}Benefit(r) = AF(c_i\xrightarrow{r}c_j.P_j)\times JS(c_i, c_j) \\
\hspace{-10pt}Cost(r)=\left\{\begin{matrix}
\sum_{p\in c_j.P_j}|c_j|\times p.type \; + & \\ \vspace{5pt} \sum_{r\in (c_j.R_j \backslash R_{ih})}|r|, & \text{if } \theta_1 <JS(c_i, c_j)\\ 
\sum_{p\in c_i.P_i}|c_i|\times p.type \; + & \\ \sum_{r\in (c_i.R_i \backslash R_{ih})}|r|, & \text{if } JS(c_i, c_j)< \theta_2,
\end{matrix}\right.
\end{array}
\end{equation}
\noindent where $JS$($c_i$, $c_j$) denotes the Jaccard similarity between $c_i.P_i$ and $c_j.P_j$, $p.type$ indicates the data type size of $p$ (e.g., the size of INT, DOUBLE, STRING, etc.), $\sum_{p\in c_j.P_j}|c_j|$$\times$ $p.type$ ($\sum_{p\in c_i.P_i}$ $|c_i|$ $\times$ $p.type$) denotes the space overheads incurred by propagating $c_j.P_j$ ($c_i.P_i$) to $c_i$ ($c_j$), 
and $\sum_{r\in (c_i.R_i \backslash R_{ih})}|r|$ ($\sum_{r\in (c_j.R_j \backslash}$ $_{R_{ih})}$ $|r|$) denotes the space overhead incurred by connecting the neighbors of $c_i$ ($c_j$) to $c_j$ ($c_i$). 

Similarly, the cost-benefit model for one-to-many rule, leveraging both data characteristics and access frequency information, is described as: 
\begin{equation}
\label{eq:bc:1m}
\begin{array}{ll}
Benefit(r) = AF(c_i\xrightarrow{r}c_j.p) \\
Cost(r) = |r|\times p.type,
\end{array}
\end{equation}
\noindent where $|r|\times p.type$ denotes the space overhead incurred by replicating $p$ as a data property of type $LIST$ to $c_i$.

As described in Section~\ref{sec:rules}, each \textit{M:N} relationship is equivalent to two {\it 1:M} relationships. Thus, we first convert each \textit{M:N} relationship in the ontology into two {\it 1:M} relationships, and then use Equation~\ref{eq:bc:1m} to decide the cost-benefit for each of them. Potentially some of the original \textit{M:N} relationships could be optimized for only one direction. This increases the flexibility of applying many-to-many rule such that more frequently accessed data properties can be propagated to the other end of the relationship.

With the cost and benefit scores, our goal is to select a subset of relationships in the ontology that maximize the total benefit within the given space limit. We map our \textbf{relationship selection problem} to the \textbf{0/1 Knapsack Problem}, which is NP-hard~\cite{Vazirani:2001}.

\begin{proposition}[Reduction]
If both benefit and cost of a relationship are positive, then every instance of the relationship selection problem can be reduced to a valid instance of the 0/1 Knapsack problem.
\label{lemma:knapsack}
\end{proposition}

The proof of Proposition~\ref{lemma:knapsack} can be found in Appendix~\ref{sec:appendix:1}. 

We adopt the fully polynomial time approximation scheme (FPTAS)~\cite{Vazirani:2001} for our relation selection problem, which guarantees that the benefit of the optimized property graph schema $Benefit(\mathcal{PGS})$ is within 1-$\epsilon$ ($\epsilon$ > 0) bound to the benefit of the optimal property graph schema $Benefit(\mathcal{PGS}_{opt})$.

\begin{algorithm}[h]
\caption{Relation-Centric Algorithm}
\label{algo:o2p:rc}
\begin{algorithmic}[1]
\Require $\mathcal{O} = (C, R, P)$, space limit $S$
\Ensure A property graph schema $\mathcal{PGS}\newline$
\hspace{90pt}\Comment{Compute Jaccard similarity for each inheritance relationship and get all parent concepts}
\ForAll {$r \in R$ of type $inheritance$}
    \State $r.js \leftarrow$ computeJS($r$)
\EndFor
\State $Benefit, Cost \leftarrow \emptyset$
\ForAll {$r_i \in R$}
    \State $Benefit[i] \leftarrow$ Benefit($r_i$)
    \State $Cost[i] \leftarrow$ Cost($r_i$)
\EndFor
\State $R_{opt} \leftarrow knapsack(R, Benefit, Cost, S)$
\ForAll {$r_i \in R_{opt}$}
    \State $O \leftarrow$ applyRules($r_i$)
\EndFor
\State $PGS \leftarrow$ generatePGS($O$)
\State \Return $PGS$
\end{algorithmic}
\end{algorithm}

Algorithm~\ref{algo:o2p:rc} takes as inputs an ontology and the space limit. Similar to Algorithm~\ref{algo:o2p:nospace}, it computes the Jaccard similarity scores for all inheritance relationships (Lines 1-2). Then it computes the cost and benefit for each relationship in the ontology $\mathcal{O}$ using Equations~\ref{eq:bc:union},~\ref{eq:bc:ih}, and~\ref{eq:bc:1m} (Lines 3-6). Next, the FPTAS algorithm is used to select the near-optimal subset of relationships $R_{opt}$ with the given space limit $S$ (Line 7). In {\it applyRules} procedure, the algorithm applies the corresponding rules; $r\in R_{opt}$ (Lines 8-9). Lastly, an optimized property graph schema is generated (Lines 10-11).

{\bf Complexity Analysis.}
The FPTAS \textit{knapsack} is the dominant procedure in Algorithm~\ref{algo:o2p:rc}, and its time complexity is $O(|R|^2\left \lfloor |R|/\epsilon \right \rfloor)$~\cite{Vazirani:2001}, where $|R|$ is the number of relationships and $\epsilon$$\in$$(0,1]$. The rest of Algorithm~\ref{algo:o2p:rc} is linear to $|R|$. Thus, the time complexity of Algorithm~\ref{algo:o2p:rc} is $O(|R|^3)$.

%% file: exp.tex
\section{Experimental Evaluation}
\label{sec:exp}

In this section, we present experiments to evaluate the effectiveness of our property graph schema design algorithms, and compare the query performance of different property graphs generated by different algorithms.

\begin{figure*}[t]
\begin{minipage}{0.49\textwidth}
    \centering
    \subfigure[Uniform Workload]{
        \includegraphics[width=0.49\columnwidth]{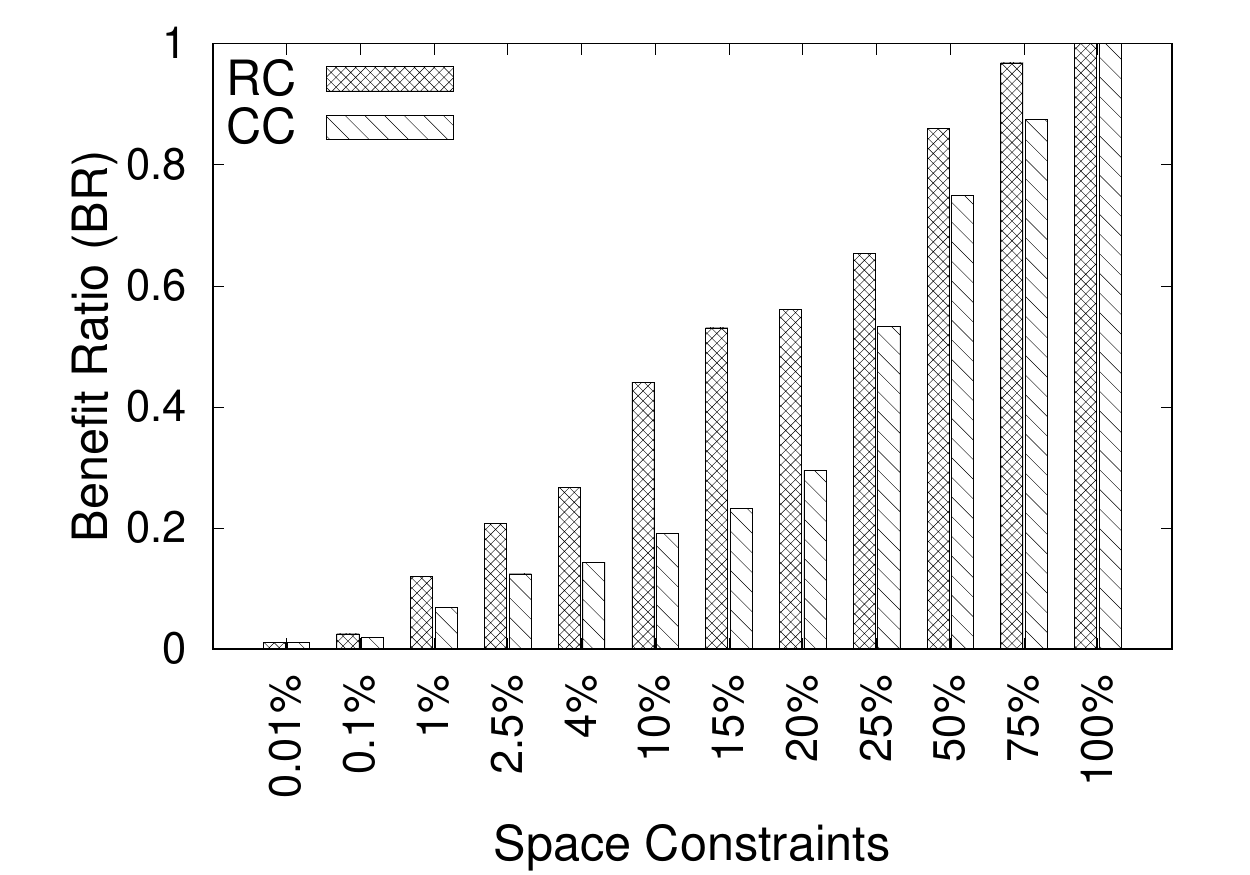}
         \label{fig:mdx:uniform}
    }\hspace{-15pt}
    \subfigure[Zipf Workload]{
        \includegraphics[width=0.49\columnwidth]{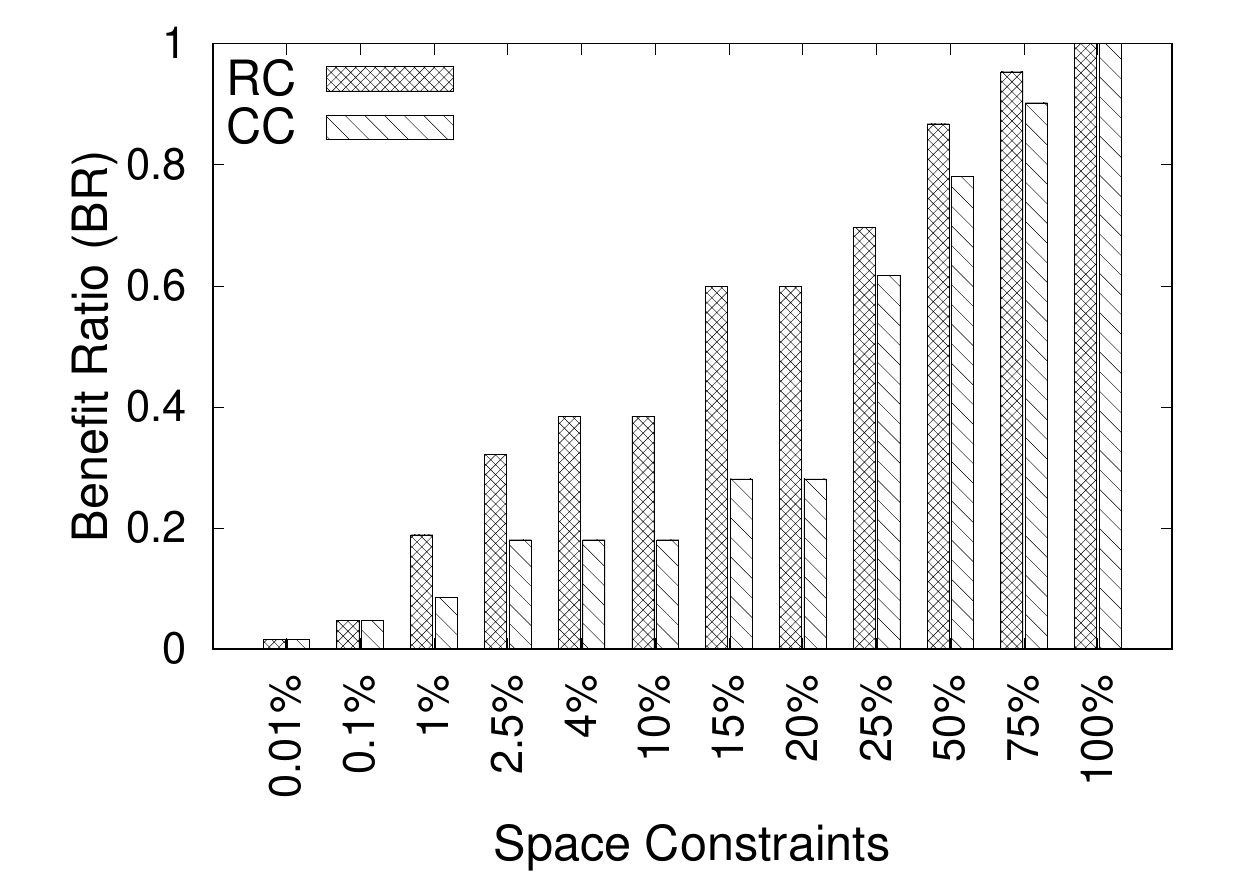}
        \label{fig:mdx:zipf}
    }
    \caption{Varying Space Constraints (MED).}
    \label{fig:mdx}
\end{minipage}
\begin{minipage}{0.49\textwidth}
    \centering
    \subfigure[Uniform Workload]{
        \includegraphics[width=0.49\columnwidth]{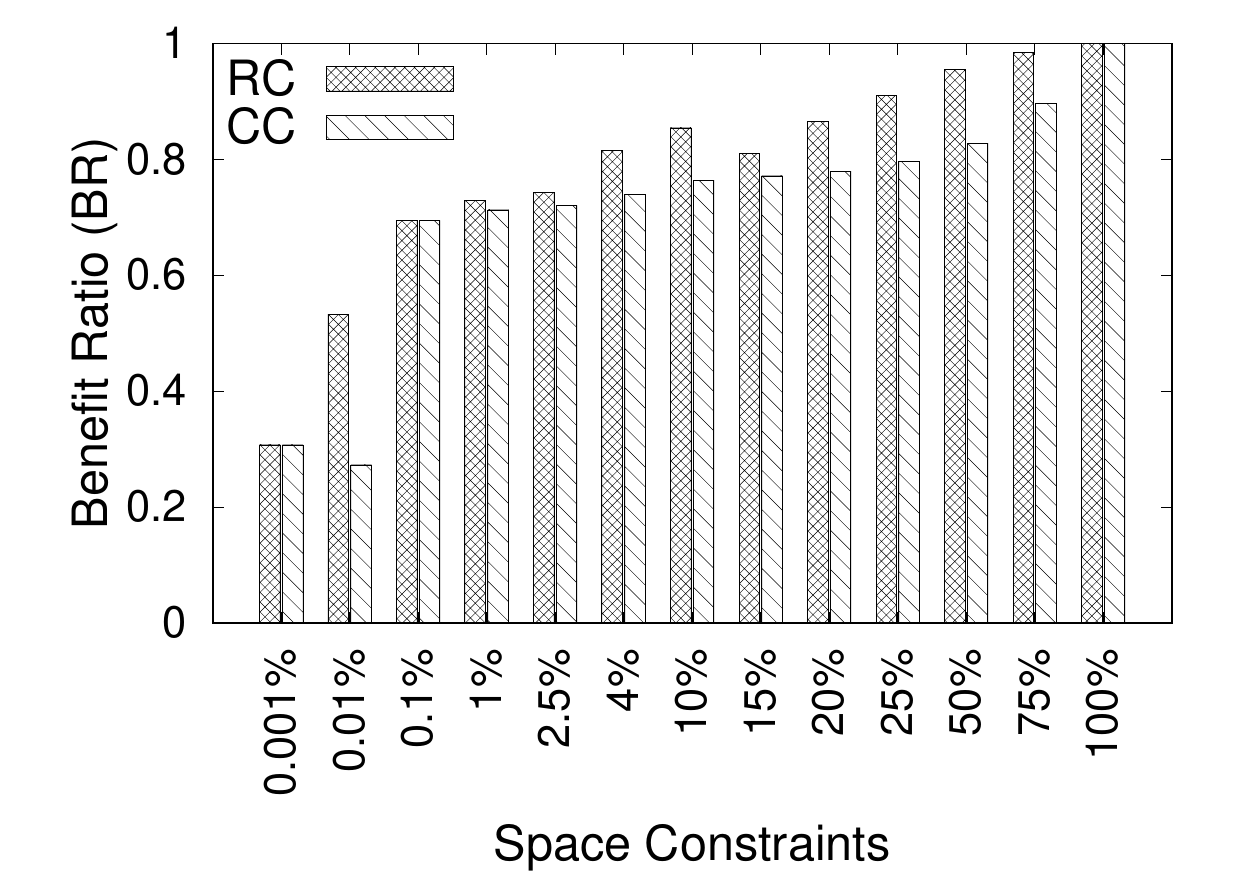}
         \label{fig:fibo:uniform}
    }\hspace{-15pt}
    \subfigure[Zipf Workload]{
        \includegraphics[width=0.49\columnwidth]{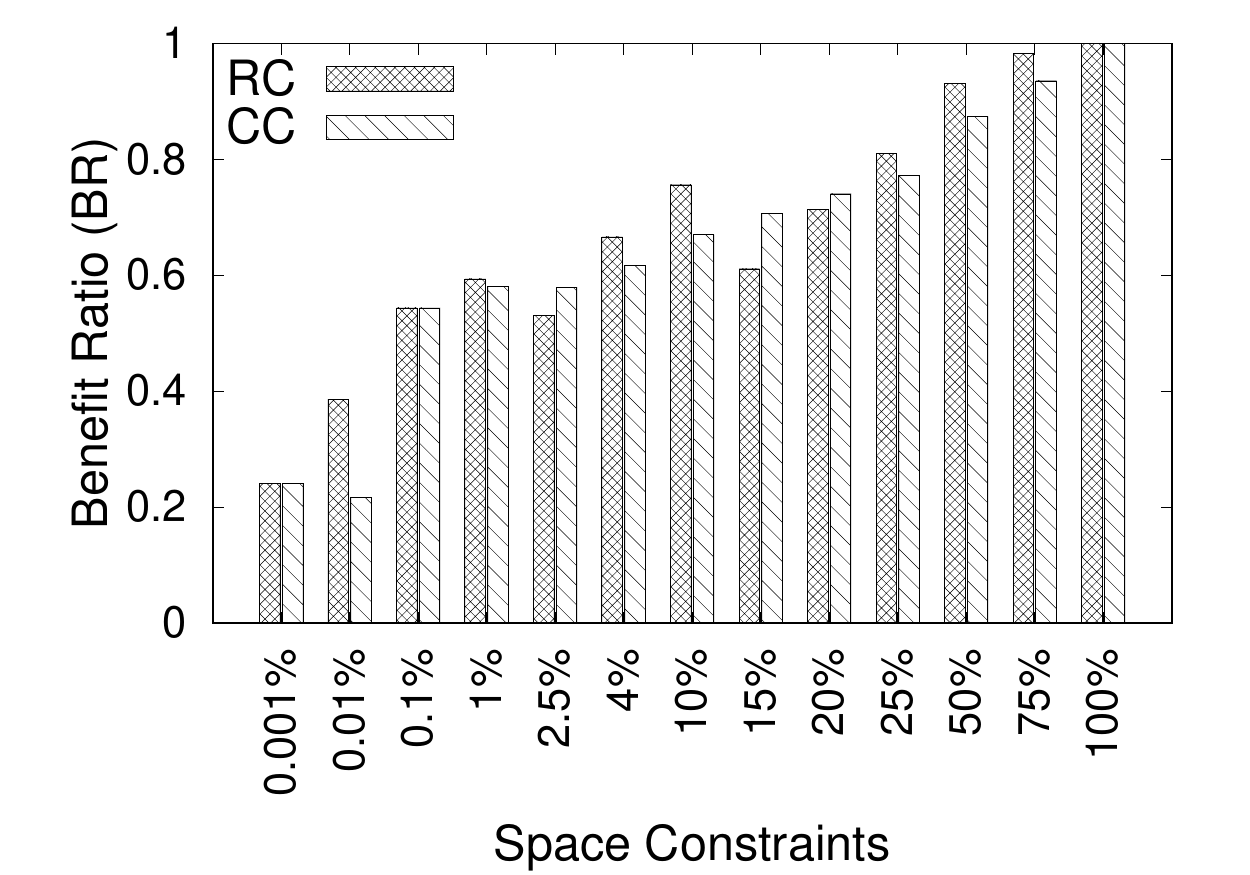}
        \label{fig:fibo:zipf}
    }
    \caption{Varying Space Constraints (FIN).}
    \label{fig:fibo}
\end{minipage}
\end{figure*}

\subsection{Experimental Setup}
\label{sec:exp:setup}

{\bf Infrastructure.} We implemented our approach in Java with JDK 1.8.0 running on Ubuntu 14.04 with 16-core 3.4 GHz CPU and 128 GB of RAM. We choose two popular graph database systems, Neo4j~\cite{neo4j} and JanusGraph~\cite{janus}, as our graph backends. We executed each experiment ten times and here we report their average.

{\bf Data sets.} To evaluate the effectiveness of our system on different application domains, we use the following two data sets and their corresponding ontologies.

1. Financial data set ({\it FIN})~\cite{blind:2018} includes data from two main sources: Securities and Exchange Commission (SEC)~\cite{SEC} and Federal Deposit Insurance Corporation~\cite{FDIC}. The size of the data set is approximately 53 GB. The corresponding financial ontology contains 28 concepts, 96 properties, and 138 relationships (4 union, 69 inheritance, and 30 one-to-many relationships). It contains financial entities, financial metrics, lender, borrower, investment relationships, the officers of the companies as well as their relationships, etc.

2. Medical data set ({\it MED}) contains medical knowledge that is used to support evidence-based clinical decision and patient education. The total size of this data set is around 12 GB. The corresponding medical ontology consists of 43 concepts, 78 properties, and 58 relationships (11 inheritance, 5 one-to-one, 30 one-to-many, and 12 many-to-many relationships).

{\bf Methodology and metrics.} To evaluate the quality of the property graph schemas produced by our algorithms, we vary the space limit and the Jaccard similarity thresholds for inheritance relationships with two different workload summaries (uniform and Zipf). Specifically, we show how effectively $PGSG$ leverages the given space limit, how robust $PGSG$ is to various workloads, and how sensitive $PGSG$ is to different similarity thresholds. $PGSG$ chooses the property graph schema with a higher total benefit score from relation-centric ($RC$) and concept-centric ($CC$) algorithms. We measure the quality of a property graph schema by $BR = \frac{B_{SC}}{B_{NSC}}$, where $B_{NSC}$ is the total benefit score of the property graph schema generated by Algorithm~\ref{algo:o2p:nospace} without any space constraint, and $B_{SC}$ indicates the total benefit score achieved by either $RC$ or $CC$ algorithm.

To verify the graph query performance, we express most graph queries in both Cypher~\cite{DBLP:conf/sigmod/FrancisGGLLMPRS18} and Gremlin~\cite{gremlin}, including path, reachability, and graph analytical queries. Among these query types, we construct a variety of query workloads conforming to different workload distributions over both financial and medical data sets. The details of these query workloads are described in Section~\ref{sec:exp:runtime}. We use latency as the metric to measure these graph queries. Latency is measured in milliseconds as the total time of all queries in a workload executed in sequential order. We also use the number of edge traversals required in a query as the second metric. It directly reveals the computational savings achieved by our optimized property graph schema. Lastly, we evaluate the efficiency of our concept-centric and relation-centric algorithms with different space constraints.

\subsection{Property Graph Schema Quality}
\label{sec:exp:opt}

{\bf Varying Space Constraint.} In Figures~\ref{fig:mdx} and~\ref{fig:fibo}, we focus on the quality of the property graph schema produced by our concept-centric ($CC$) and relation-centric ($RC$) algorithms compared to our method without space constraints $NSC$ (Algorithm~\ref{algo:o2p:nospace}). We choose two commonly seen workload summaries, uniform and Zipf distributions. The Zipf workload gives more access to the key concepts in the ontology. Namely, the access frequencies of concepts in the ontology follow either uniform or Zipf distribution. And the skew factor of Zipf distribution is set to 1. We first use $NSC$ to produce an optimal property graph schema $PGS_{NSC}$ without any space constraint, and then compute the total benefit score $B_{NSC}$ achieved by $PGS_{NSC}$ as well as the total amount of space $S_{NSC}$ needed by $PGS_{NSC}$. We also compute the total amount of space $S_{DIR}$ needed by the direct mapping algorithm from the given ontology. The space used by {\em NSC} is approximately 29GB for {\em MED} and 106GB for {\em FIN}, respectively. The total amount of space needed by the direct mapping algorithm $S_{DIR}$ is 12GB for {\em MED} and 53GB for {\em FIN}, respectively. We, then, vary the space constraint from $S_{DIR}$ to $S_{NSC}$, such that the range of the Y-axis in Figures~\ref{fig:mdx} and~\ref{fig:fibo} is from 0 to 1. Figures~\ref{fig:mdx} and~\ref{fig:fibo} show results from {\it MED} and {\it FIN} data sets respectively.

In Figure~\ref{fig:mdx}, we observe that $RC$ consistently outperforms $CC$ with both uniform and Zipf workloads. The reason is that $RC$ has a global ordering of all relationships, and the global ordering is near-optimal with respect to the given space constraint due to the adopted approximate Knapsack algorithm. On the contrary, $CC$ suffers from a rather local optimal ordering with respect to each concept. Hence, it misses the opportunity to utilize the space for more beneficial relationships. Moreover, we observe that with approximately 20\% of the maximum space constraint, {\em RC} is able to produce high-quality property graph schemas which achieve above 50\% of the total benefit. In other words, both algorithms can effectively utilize the rather limited space. Lastly, both $RC$ and $CC$ produce the same property graph schema as $PGS_{NSC}$ when the space constraint reaches 100\%, which substantiates Theorem~\ref{theo:order}.

Similarly, $RC$ outperforms $CC$ In Figure~\ref{fig:fibo}, as $CC$ utilizes the space for one concept at a time, missing the opportunities for more beneficial relationships in the ontology. We also observe that both algorithms, with uniform and Zipf workloads, have a couple of drops when the space constraint increases. The reason is primarily due to the complexity of {\it FIN} ontology. Given that the inheritance relationships are more dominant in {\it FIN}, the given space may be exhausted quickly by certain inheritance relationships. Again, $RC$ and $CC$ produce the same property graph schema as $PGS_{NSC}$ with 100\% space constraint.

\begin{figure}[!htb]
\centering
\subfigure[Uniform Workload]{
\includegraphics[width=0.49\columnwidth]{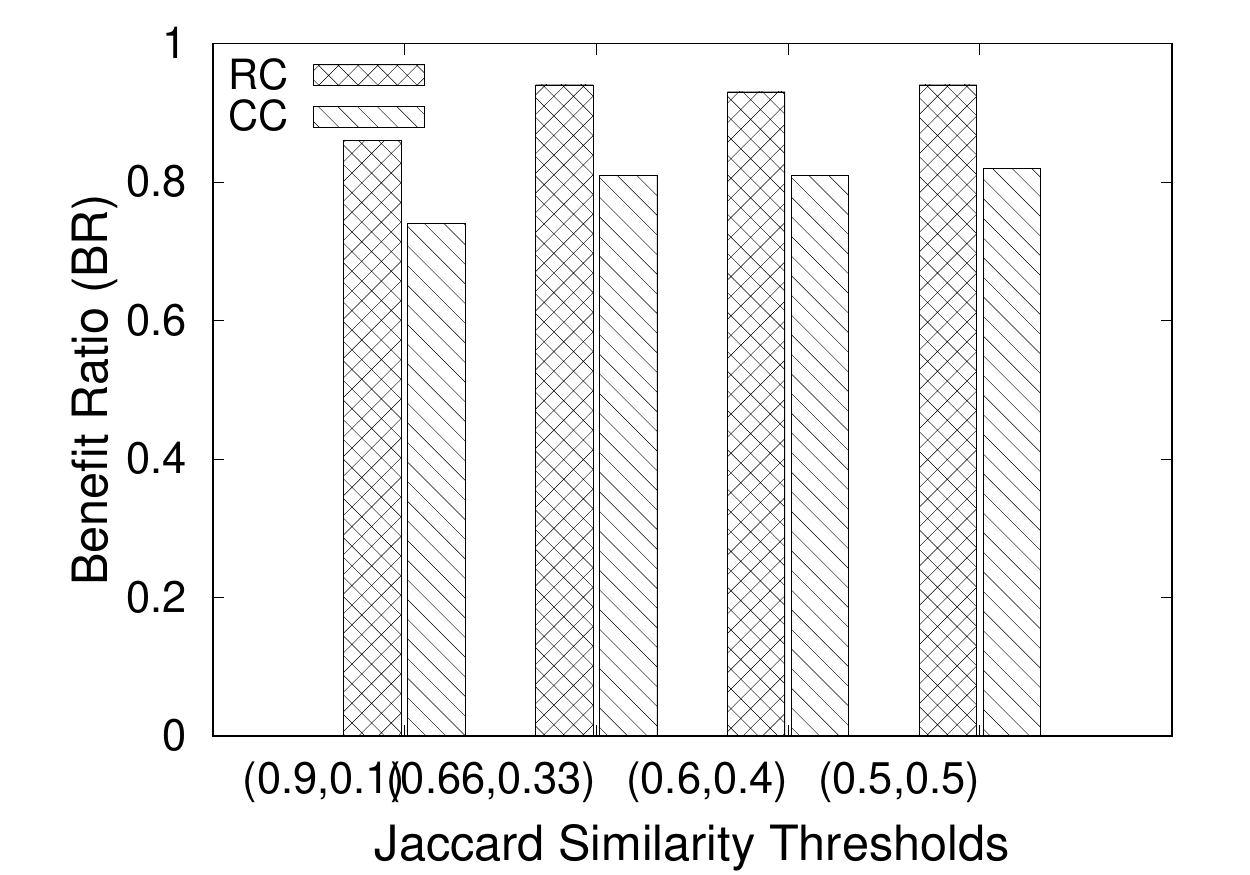}  
\label{fig:js:uniform}
}\hspace{-15pt}
\subfigure[Zipf Workload]{
\includegraphics[width=0.49\columnwidth]{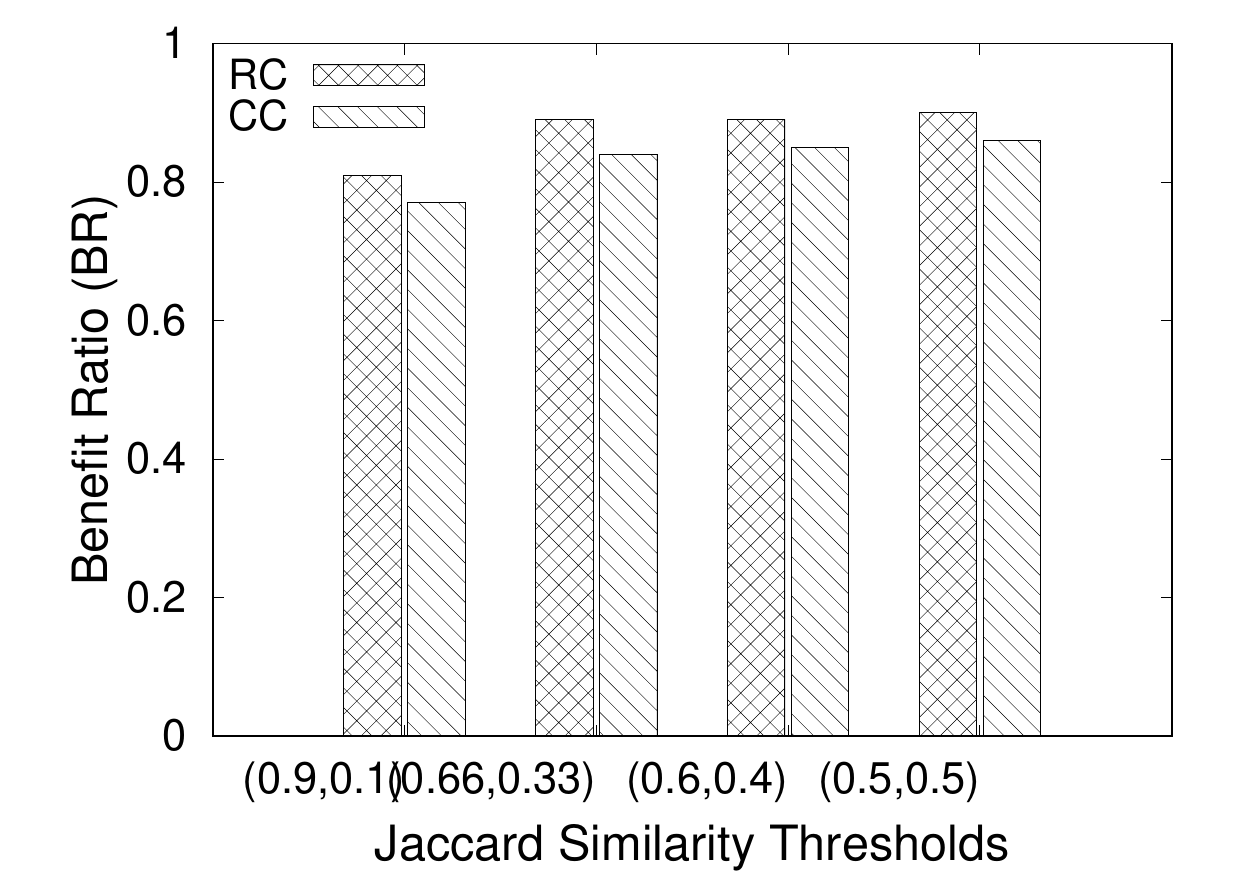}
\label{fig:js:zipf}
}
\caption{Varying Jaccard Thresholds (FIN).} 
\label{fig:js}
\end{figure}

{\bf Varying Jaccard Similarity.} In Figure~\ref{fig:js}, we show the sensitivity of both $CC$ and $RC$ with respect to the Jaccard similarity thresholds ($\theta_1$ and $\theta_2$). In this experiment, we choose {\it FIN} ontology because it consists of multiple inheritance relationships. Uniform and Zipf workload distributions are used to examine the robustness of our $CC$ and $RC$ algorithms. Note that the space constraint in this experiment is set to ($S_{NSC}$-$S_{DIR}$)/2 under each specific Jaccard similarity threshold. The reason is that the cost (space overhead) of the same inheritance relationship can vary (Eq.~\ref{eq:bc:ih}) depending on the similarity threshold. Consequently, the space consumption of the optimal property graph changes under different thresholds. As shown in Figure~\ref{fig:js}, both $CC$ and $RC$ are robust under different similarity thresholds. In the worst case, they achieve more than 70\% of the maximum benefit score under 50\% space constraint. This shows that when the cost-benefit of an inheritance relationship changes due to a different threshold, both {\em CC} and {\em RC} can adjust accordingly by choosing different and more beneficial relationships to optimize. Hence, the total benefit scores achieved by both algorithms are relatively stable.

In summary, $CC$ and $RC$ produce high quality property graph schemas under various settings. They work effectively with any given space constraints. Moreover, $RC$ always produces a near-optimal property graph schema and outperforms $CC$ in most cases. Our property graph schema generator still leverages both algorithms to choose the property graph schema with the highest benefit score under any space constraints. 

\subsection{Graph Query Execution}
\label{sec:exp:runtime}

In this section, we focus on the graph query execution performance over the property graphs created by our ontology-driven approach. We use both {\it MED} and {\it FIN} data sets to conduct our experiments. First, we create a micro benchmark to empirically examine whether the property graph schema from our approach can actually benefit a set of graph primitives including simple pattern matching, vertex property lookup, and aggregation on vertices. Second, we study the overall execution time for a given graph query workload by mixing the above graph primitives. We run the graph queries, expressed in Cypher and Gremlin, on Neo4j and JanusGraph, respectively. Note that our goal is not to compare the performance between two systems, rather to show that our schema optimization results in query performance improvements irrespective of the backend. 
\begin{figure*}[!htbp]
\begin{center}
\includegraphics[width=0.86\textwidth]{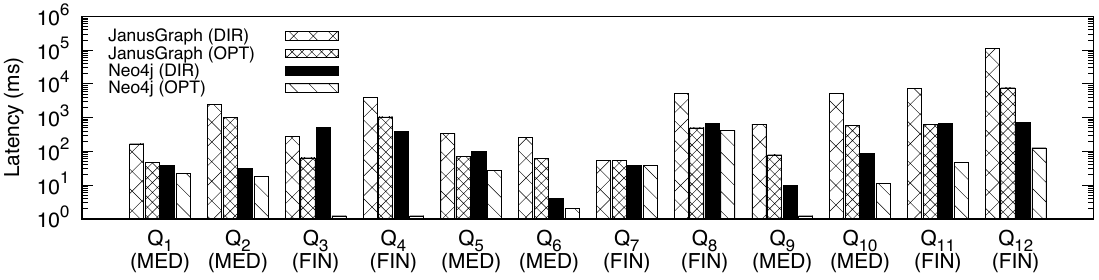}
\caption{Microbenchmark - Pattern Matching ($Q_1$-$Q_4$), Property Lookup ($Q_5$-$Q_8$), Aggregation ($Q_9$-$Q_{12}$).}
\label{fig:primitive}
\end{center}
\end{figure*}

{\bf Microbenchmark Using Graph Primitives.} With both {\it MED} and {\it FIN} data sets, we compare the query performance of the property graph created by the optimized graph schema ({\it OPT}) to the baseline property graph created by a direct mapping of the ontology ({\it DIR}). 
The following parameter settings are used to produce $OPT$: Jaccard similarity thresholds $\theta_1 = 66\%$, $\theta_2 = 33\%$, and space constraint 0.5 ($S_{NSC}-S_{DIR}$). All queries ($Q_1$-$Q_{12}$) are first expressed against {\it DIR} and then rewritten into the semantically equivalent queries over {\it OPT}. These queries are constructed according to the query patterns in~\cite{DBLP:journals/pvldb/BonifatiMT17}. We show a few representative queries used in the microbenchmark below.

\begin{lstlisting}
$Q_1$: MATCH (d:Drug)-[p:cause]->(r:Risk)<-
[p2:unionOf]-(ci:ContraIndication)
RETURN d.name
\end{lstlisting}

\begin{lstlisting}
$Q_3$: MATCH (aa:AutonomousAgent)<-[r1:isA]-
(p:Person)<-[r2:isA]-(cp:ContractParty)
RETURN aa
\end{lstlisting}

\begin{lstlisting}
$Q_5$: MATCH (dl:DrugLabInteraction)-[r:isA]->
(di:DrugInteraction)
RETURN di.summary
\end{lstlisting}

\begin{lstlisting}
$Q_7$: MATCH (n:Corporation) 
RETURN n.hasLegalName
\end{lstlisting}

\begin{lstlisting}
$Q_9$: MATCH p=(d:Drug)-[r:hasDrugRoute]->
(dr:DrugRoute)
RETURN dr.drugRouteId, size(COLLECT(
d.brand)) AS numberOfDrugBrands
\end{lstlisting}

\begin{lstlisting}
$Q_{11}$: MATCH p=(con:Contract)-[r:isManagedBy]->
(corp:Corporation)
RETURN size(COLLECT(con.hasEffectiveDate)) AS numberOfEffectiveDates
\end{lstlisting}

As shown in Figure~\ref{fig:primitive}, the results are unequivocal. The optimized schema has significant advantages over the direct mapping schema for all types of queries. The graph pattern matching queries ($Q_1$-$Q_4$) report all matches of a sub-graph with 3 vertices and 2 edges in the property graph. Query execution times with our approach are at least 2.4 times faster than the direct mapping schema. The number of edge traversals on {\it DIR} is always 2 as the query is specified with 2 edges connecting 3 vertices. On the other hand, our property graph only requires at most 1 edge traversal as some of the neighbor vertices have been already merged with the starting vertices.

$Q_5$-$Q_8$ are vertex property lookup queries. Both $Q_5$ and $Q_8$ are interested in a property of a vertex of a parent concept, and the starting vertex is a vertex of a child concept. $Q_6$ starts from a vertex and looks for a property of its neighbor vertex. {\it OPT} has the property of type $List$ with the starting vertex, and is able to return the result without any edge traversal. $Q_7$ looks for a property of the starting vertex. In this case, {\it OPT} and {\it DIR} have identical query performance as no edge traversal is required. Hence {\it OPT} takes advantage of having the property of the parent concept available at the starting vertex, and consequently returns the result without any edge traversals. Therefore, the query runs more than an order of magnitude slower on the property graph of {\it DIR} than the one on {\it OPT} in the worst case.

$Q_9$-$Q_{12}$ are graph aggregation queries that involve traversal from one vertex to the other. They count the number of neighbors of the starting vertex. On average, the query execution time is an order of magnitude faster for {\it OPT} approach compared to {\it DIR}. Again, the reason is that the aggregation on the neighbor vertices can be instantaneously returned from the starting vertex. The above results suggest that using the proposed ontology-driven approach can bring significant benefits to a variety of graph queries.

Lastly, we observe that the performance gain on Neo4j is more substantial compared to JanusGraph (e.g., $Q_3$, $Q_4$, $Q_9$, etc.). This shows that disk-based graph systems (e.g., Neo4j) benefits much more from our techniques, as the optimized schema requires significantly less disk I/O. Namely, the graph system loads less number of vertices and edges into memory. We expect such benefit to become even greater when the size of the property graph increases.

\begin{table}[!htb]
\small
\centering
\caption{Microbenchmark - \# Edge Traversals.}
\begin{tabular}{|c|p{1cm}|p{1cm}||c|p{1.1cm}|p{1cm}|}
\hline
\textit{Graph} & \multicolumn{2}{c||}{$\#$ \textit{Edge Traversals}} & \textit{Graph} & \multicolumn{2}{c|}{$\#$ \textit{Edge Traversals}} \\ \cline{2-3} \cline{5-6}
\textit{Queries} & \textit{DIR} & \textit{OPT} & \textit{Queries} & \textit{DIR} & \textit{OPT} \\ \hline\hline
$Q_1$     & 21,608   & 6,072  & $Q_7$    & 0       & 0 \\ \hline
$Q_2$     & 288,142  & 115,014& $Q_8$    & 493,588 & 0 \\ \hline
$Q_3$     & 36,272   & 0      & $Q_9$    & 67,397  & 0 \\ \hline
$Q_4$     & 510,460  & 97,614 & $Q_{10}$ & 429,636 & 15,327 \\ \hline
$Q_5$     & 38,768   & 0      & $Q_{11}$ & 524,265 & 0     \\ \hline
$Q_6$     & 32,586   & 0      & $Q_{12}$ & 110,4756& 548,262\\ \hline
\end{tabular}
\label{tab:primitive}
\end{table}

In addition, Table~\ref{tab:primitive} reveals that {\em OPT} substantially reduces the number of edge traversals required in most queries, which leads to significant computational savings and performance gains. In several cases (e.g., $Q_3$, $Q_6$), edge traversals can be completely avoided as the queried information is available locally within the starting vertices. On the other hand, the performance gains of certain queries (e.g., $Q_5$, $Q_8$, $Q_{12}$) are not as significant as others, even though the number of edge traversals with {\em OPT} is much smaller than the one with {\em DIR}. The reason is that the costs of lookup and return operations are non-trivial in both {\em DIR} and {\em OPT}, which can be observed from the latency of these queries in Fig.~\ref{fig:primitive}.

{\bf Graph Query Workload Performance.} To evaluate the runtime performance of the property graph schema generated by our approach, we first generate two query workloads, including both uniform and Zipf distributions in terms of the access frequency of the concepts in the ontology. We vary the Zipf's skew factor from 0 (i.e., uniform distribution) to 2 (highly skewed). All query workloads consist of 15 queries of mixed types (i.e., pattern matching, lookup, and aggregation), similar to the ones used in the microbenchmark. The space limit is set to 20\% of the space consumed by {\em NSC} (i.e., 15.4GB for {\em MED} and 80GB for {\em FIN}). The Jaccard similarity thresholds are $\theta_1 = 66\%$ and $\theta_2 = 33\%$. The optimized schemas (\textit{OPT$_{\textit{MED}}$} and \textit{OPT$_{\textit{FIN}}$}) are produced by the best performing algorithm of {\em RC} and {\em CC}.

\begin{table}[!htb]
\small
\centering
\caption{Benefit Ratio w.r.t {\em B$_{NSC}$}.}
\begin{tabular}{|c||c|c|c|c|c|c|c|c|}
\hline
\textit{Skew} & \multicolumn{4}{c|}{\textit{MED}} & \multicolumn{4}{c|}{\textit{FIN}} \\ \cline{2-5} \cline{6-9}
\textit{Factor} & 0 & 1 & 1.5 & 2 & 0 & 1 & 1.5 & 2 \\ \hline\hline
{\em RC} & 56\% & 59\% & 62\% & 71\% & 67\% & 71\% & 74\% & 88\% \\ \hline
{\em CC} & 30\% & 43\% & 50\% & 63\% & 65\% & 74\% & 80\% & 88\% \\ \hline
\end{tabular}
\label{tab:workload:br}
\end{table}

Table~\ref{tab:workload:br} shows the quality of the property graph schema produced by {\em RC} and {\em CC} compared to the one without space constraints {\em NSC}. We observe that both {\em RC} and {\em CC} correctly prioritize the most cost-effective relationships when the workloads are highly skewed. {\em RC} performs better than {\em CC} over {\em MED}, because {\em MED} has more data properties per concepts and {\em RC} makes more flexible decisions in terms of which relationships to optimize. On the other hand, {\em CC} performs better than {\em RC} over {\em FIN} as it successfully selects few concepts that are frequently accessed by the highly skewed workloads.

\begin{figure}[!htb]
\centering
\subfigure[JanusGraph]{
\includegraphics[width=0.49\columnwidth]{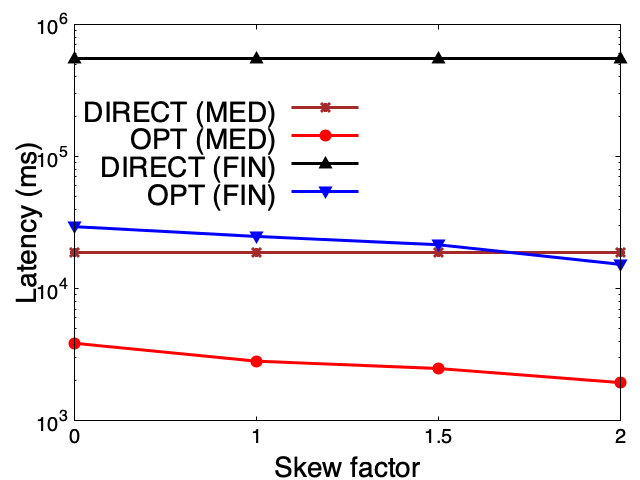}  
\label{fig:workload:janus}
}\hspace{-15pt}
\subfigure[Neo4j]{
\includegraphics[width=0.49\columnwidth]{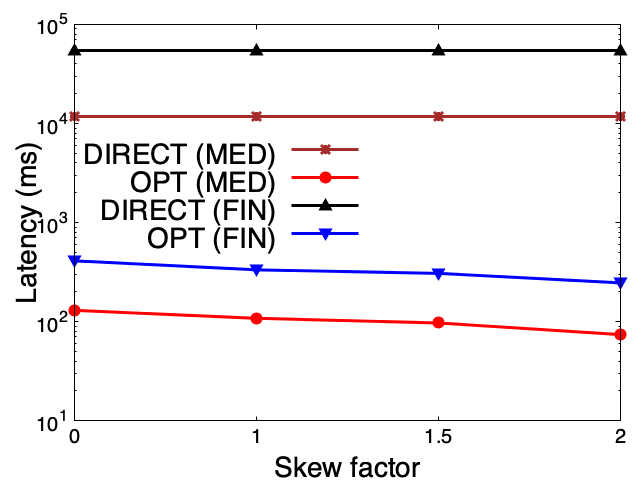}
\label{fig:workload:neo4j}
}
\caption{Total Query Latency (MED \& FIN).} 
\label{fig:workload}
\end{figure}

We compare our optimized schemas to the direct mapping schemas (\textit{DIRECT$_{\textit{MED}}$}, \textit{DIRECT$_{\textit{FIN}}$}) on both JanusGraph and Neo4j. The total query latency measures the performance on these property graphs corresponding to different schemas. Fig.~\ref{fig:workload} shows the total query latency in log scale. Both \textit{OPT$_{\textit{MED}}$} and \textit{OPT$_{\textit{FIN}}$} offer significant performance boosts to the graph query workloads on both JanusGraph and Neo4j. In Fig.~\ref{fig:workload:janus}, we observe that the total query latency on the optimized schema, on average, is around 7 and 26 times faster than the direct mapping one over {\it MED} and {\it FIN}, respectively. The winning margin is substantially bigger (i.e., 129 and 176 times faster) on Neo4j (Fig.~\ref{fig:workload:neo4j}). The total query latency on both optimized schema is approximately 2 orders of magnitude faster than the direct mapping. Moreover, we also observe that the total query latency decreases with increasing skew factor. Both \textit{OPT$_{\textit{MED}}$} and \textit{OPT$_{\textit{FIN}}$} achieve the lowest latency when the workload distributions are highly skewed. This indicates that the most frequently accessed concepts and relationships in the workloads are chosen to be optimized given the space limit. Based on these results, we verify that the designed rules for different types of relationships in the ontology are effective in terms of reducing edge traversals and consequently improving the graph query performance. Furthermore, we demonstrate that our approach can effectively utilize the given space constraint by leveraging data distribution and workload summaries.

\subsection{Efficiency of Property Graph Schema Algorithms}
\label{sec:exp:efficiency}

Finally, we study the execution time of our concept-centric and relation-centric algorithms (Table~\ref{table:opt}). First, we observe that both \textit{CC} and \textit{RC} produce an optimized property graph schema in less than one second with different space constraints (shown in Table~\ref{table:opt} as percentages of the space consumed by Algorithm~\ref{algo:o2p:nospace}). The optimization time of both algorithms is negligible compared to an exhaustive search approach, which even failed to produce an optimal schema for \textit{MED} after 3 hours. Second, neither of the algorithms is sensitive to the space constraint, since both algorithms have a polynomial time complexity with respect to the number of concepts and relationships in the given ontology. Third, \textit{RC} is consistently faster than \textit{CC}, and the performance difference is more significant in \textit{FIN}. This is due to the cost of \textit{ontologyPR} procedure being dominant in \textit{CC}. It usually takes more iterations to converge when the ontology (i.e., \textit{FIN}) is more complex.

\begin{table}[!htb]
\centering
\caption{Efficiency of \textit{RC} $\&$ \textit{CC} (Time in \textit{ms}).}
\begin{tabular}{| c | c | c | c | c | c | c |}
\hline
                          & \multicolumn{3}{c|}{\textit{MED}} &\multicolumn{3}{c|}{\textit{FIN}} \\ \hline
\textit{Space Constraint} & \textit{25\%} & \textit{50\%} & \textit{75\%} & \textit{25\%} & \textit{50\%} & \textit{75\%}\\ \hline
\textit{RC}        & 23 & 23 & 26 & 192 & 188 & 193 \\ \hline
\textit{CC}        & 34 & 36 & 36 & 373 & 344 & 372 \\ \hline
\end{tabular}
\label{table:opt}
\end{table}

%% file: related_work.tex
\section{Related Work}
\label{sec:related}

Schema optimization for improving query performance has been studied in the database community for decades~\cite{Codd:1970,Finkelstein:1988,Zilio:2004,7498239}. In recent years, the emergence of many large-scale knowledge graphs has drawn attention for schema optimization. In this section, we present important works in this field, highlighting the main differences to our approach.

\textbf{Schema Optimization in RDBMS/NoSQL.} Extensive work is available for schema design problem in relational database systems~\cite{Finkelstein:1988,Agrawal:2000,Zilio:2004,Bruno:2005,Kimura:2010,Dash:2011}. RDBMSs provide a clean separation between logical and physical schemas. The logical schema includes a set of table definitions and determines a physical schema consisting of a set of base tables~\cite{Finkelstein:1988,Agrawal:2000,Zilio:2004}. The physical layout of these base tables is then optimized with auxiliary data structures such as indexes and materialized views for the expected workload~\cite{Agrawal:2000,Kimura:2010}. Typically, the physical design often involves identifying candidate physical structures and selects a good subset of these candidates~\cite{Dash:2011}. NoSE~\cite{7498239} is introduced to recommend schemas for NoSQL applications. Its cost-based approach utilizes a binary integer programming formulation to generate a schema based on the conceptual data model from the application.

In principle, our approach is similar to the logical schema design in RDBMSs, which defers the physical design to the underlying graph systems. Other than that, our approach is different from the above methods since the data modeling for graphs is inherently different from the relational data model. Specifically, the graph structure results in more expressive data models than those produced using relational databases, allowing the formation of graph queries (e.g., reachability, path finding, pattern matching) in a very intuitive fashion. Moreover, our approach exploits the rich semantic information available in an ontology to drive the schema optimization, which is not considered by any of the previous works.

\textbf{Schema Optimization in Knowledge Graphs.} 
In the last few years, RDF has been growing significantly for expressing graph data. A variety of schemas have been proposed for physically storing graph data in both centralized and distributed settings~\cite{Huang2011ScalableSQ,DBLP:conf/www/MadukoASS07,DBLP:journals/vldb/NeumannW10,DBLP:conf/icde/NeumannM11,DBLP:conf/icde/MeimarisPMA17,DBLP:conf/sigmod/BorneaDKSDUB13,DBLP:conf/wise/HarrisS05,DBLP:journals/vldb/AbadiMMH09,DBLP:conf/vldb/ChongDES05}. Some of these works focus on optimizing RDF data storage and SPARQL queries based on either workload statistics~\cite{DBLP:conf/www/MadukoASS07,DBLP:journals/vldb/NeumannW10,DBLP:conf/icde/NeumannM11,DBLP:conf/icde/MeimarisPMA17} or heuristics~\cite{DBLP:conf/edbt/TsialiamanisSFCB12}. A fundamental difference to those works is that we neither re-load the data to follow a new schema, nor build new indices, nor optimize the queries on-the-fly (e.g., join reordering). Instead, inspired by database literature as stated above, we provide an optimized property graph schema design before loading the data, and directly instantiate a property graph conforming to this schema on a graph database. Graph queries are then executed on the graph database, where graph query optimization techniques can be further utilized. Other works~\cite{DBLP:conf/sigmod/BorneaDKSDUB13,DBLP:conf/wise/HarrisS05,DBLP:journals/vldb/AbadiMMH09,DBLP:conf/vldb/ChongDES05} attempt to transform RDF data into relational data and provide SPARQL views over relational schemas, leveraging the many years of experience in RDBMS schema optimization. Unlike those approaches, we do not create views over a property graph. Instead, we directly express property graph queries in Cypher or Gremlin over optimized property graphs. Whether a graph database internally uses views or not for query optimization is orthogonal to this work.

Recently, works such as~\cite{DBLP:conf/sigmod/SunFSKHX15,szarnyas2017incremental,DBLP:conf/edbt/HassanKJAS18} address a similar problem in the context of property graphs. GRFusion~\cite{DBLP:conf/edbt/HassanKJAS18} focuses on filling the gap between the relational and the graph models rather than optimizing the graph schema to achieve better query performance. Sz{\'a}rnyas et al.~\cite{szarnyas2017incremental} propose to use incremental view maintenance for property graph queries. However, their approach can only support a subset of property graph queries by using nested relational algebra. SQLGraph~\cite{DBLP:conf/sigmod/SunFSKHX15} and Db2 Graph~\cite{DBLP:conf/sigmod/TianXZPTSKAP20} introduce a physical schema design that combines relational storage for adjacency information with JSON storage for vertex and edge attributes. It also translates Gremlin queries into SQL queries in order to leverage relational query optimizers. However SQLGraph and Db2 Graph also focus on physical schema design which only targets on the relational databases. The query translator is limited to Gremlin queries with no side effects. Our ontology-driven approach is different for the following reasons. First, our approach produces a high-quality schema applicable to any graph system compatible with property graph model and Gremlin or Cypher queries. Second, we exploit the rich semantic information in an ontology to guide the schema design. Last but not least, our approach can further leverage these techniques to decide how the property graph should be stored on different storage backends.

Materialized views~\cite{DBLP:journals/tkde/FanWW16,DBLP:conf/icde/TrindadeKCMS20} are also introduced to answer graph pattern queries. Views are either given as inputs or generated based on query workloads. Then a subset of views are chosen to answer a query. Hence the optimized schema generated from our approach can be considered as a view on the original property graph, which can be consumed by their technique.

%% file: conclusions.tex
\section{Conclusions}
\label{sec:conclusion}

To the best of our knowledge, our ontology-driven approach is the first to address the property graph schema optimization problem for domain-specific knowledge graphs. Our approach takes advantages of the rich semantic information in an ontology to drive the property graph schema optimization. The produced schemas gain up to 3 orders of magnitude graph query performance speed-up compared to a direct mapping approach in two real-world knowledge graphs.

%% file: appendix1.tex
\section{Proof Sketch of Theorem~\ref{theo:order}}
\label{sec:appendix:2}

\begin{proof}
Let $O$ = ($C$, $R$, $P$) be an ontology given as input to Algorithm~\ref{algo:o2p:nospace}, 
and let $O_{out}$ = ($C_{out}$, $R_{out}$, $P_{out}$) be the resulting ontology, which is used in Line 18 to produce the output $\mathcal{PGS}$. 
Proving Theorem~\ref{theo:order} is equivalent to proving that applying the rules for any $R' \subseteq R$ in any order will yield the same result $O_{out}$. 
The theorem trivially holds when $|R'| = 0$ ($O_{out} = O$), and when $|R'| = 1$ (only one rule can be triggered). 

{\bf Base case.} $|R'| = 2$, i.e., for any two relationships, applying the rules in any order yields the same result. Since we only have two relationships, only two rules will be triggered if the relationships are of different types, or one rule will be triggered twice if the two relationships are of the same type. Therefore, we need to prove that applying each pair of rules in any order will yield the same results, examining every possible scenario for each rule. 

Specifically, we need to prove that the following pairs of rules are order-independent:
($i$) union rule and inheritance rule, ($ii$) inheritance rule and 1:$M$ rule, ($iii$) union rule and 1:$M$ rule, ($iv$) inheritance rule and $M$:$N$ rule, ($v$) union rule and $M$:$N$ rule, and ($vi$) 1:$M$ rule and $M$:$N$ rule. 

{\em (i) Union and Inheritance.} To prove that union and inheritance rules are order-independent, we examine all the cases in which those two rules may be triggered in the same graph, as shown in Figure~\ref{fig:union_and_inheritance}(a), (b), (c). We assume that the Jaccard similarity between the two concepts connected with an inheritance relationship is less than $\theta_2$ (see Algorithm~\ref{algo:isa:nospace}), so the inheritance rule is triggered and the properties of the parent concept are copied to the child concept. It is straightforward to apply the following observations to the case in which the Jaccard similarity is greater than $\theta_1$ as well. Figure~\ref{fig:union_and_inheritance} contains more than two relationships, but only two relationships are sufficient to prove the case\footnote{Consider only the relationships ($c_1$, $c_2$), ($c_6$, $c_5$) for Figure~\ref{fig:union_and_inheritance}(a), ($c_1$, $c_2$), ($c_1$, $c_5$) for Figure~\ref{fig:union_and_inheritance}(b), and ($c_1$, $c_2$), ($c_5$, $c_2$) for Figure~\ref{fig:union_and_inheritance}(c).}. The additional relationships shown are for illustration purpose only. 

\begin{figure}[htb]
\centering
\includegraphics[width=1\columnwidth]{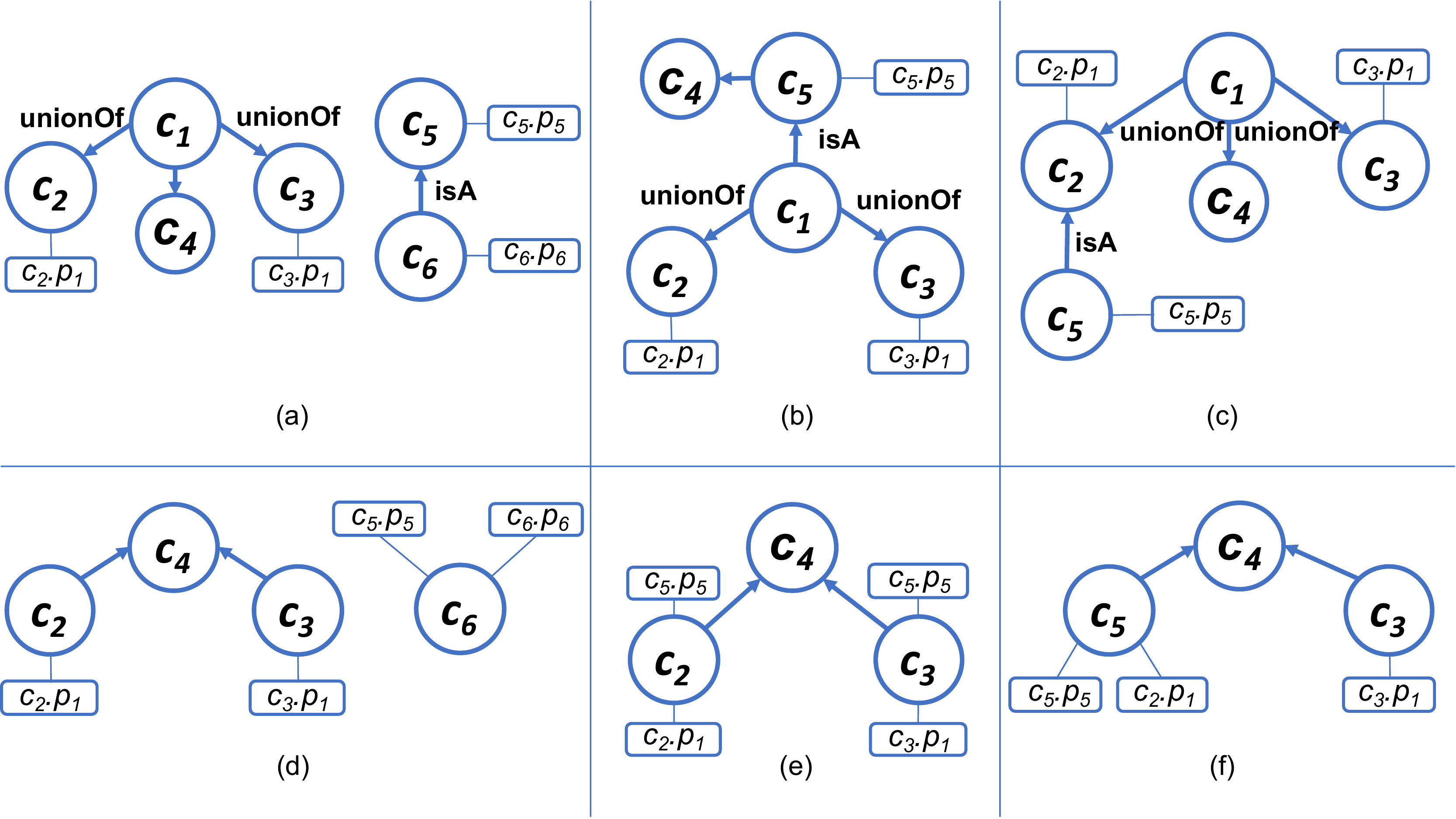}  
\caption{Union and Inheritance Rules Independence.}
\label{fig:union_and_inheritance}
\end{figure} 

In the trivial case of Figure~\ref{fig:union_and_inheritance}(a), the source and destination concepts of the union and inheritance relationships are not inter-connected. 
If we apply the union rule first, we will end up with the left part of Figure~\ref{fig:union_and_inheritance}(d), leaving the right part of Figure~\ref{fig:union_and_inheritance}(a) unchanged, and 
if we apply the inheritance rule first, we end up with the right part of Figure~\ref{fig:union_and_inheritance}(d), leaving the left part of Figure~\ref{fig:union_and_inheritance}(a) unchanged. In both cases, applying the second rule generates the graph of Figure~\ref{fig:union_and_inheritance}(d). 

The case shown in Figure~\ref{fig:union_and_inheritance}(b) is more complex, where the same concept ($c_1$) corresponds to a union concept and a child concept. 
Applying the union rule first, we remove $c_1$ and connect its member concepts $c_2$ and $c_3$ to $c_5$ through inheritance relationships. Note that those inheritance relationships come with the same Jaccard value as the original one connecting $c_1$ to $c_5$, which we have assumed to be less than $\theta_2$. Then, the inheritance rule is triggered, removing $c_5$, copying its properties to its new children $c_2$ and $c_3$, and connecting them to $c_4$, as shown in Figure~\ref{fig:union_and_inheritance}(e). 
If we apply inheritance first, instead of union, then we first remove $c_5$, copy its properties to $c_1$ and connect $c_1$ to $c_4$. Then, applying the union rule, we remove $c_1$ and connect the member concepts $c_2$ and $c_3$ to $c_4$, again resulting in the graph of Figure~\ref{fig:union_and_inheritance}(e). 
The same observations hold for the case in which $c_1$ corresponds to a parent concept and a union concept. 

In a similar way, we can show that union and inheritance rules are order-independent in the case of Figure~\ref{fig:union_and_inheritance}(c), in which the same concept ($c_2$) corresponds to a member concept and a parent concept. 
If we apply the union rule first, we remove $c_1$ and connect the member concepts $c_2$ and $c_3$ to $c_4$. Then, applying the inheritance rule, we remove $c_2$, copy its properties to $c_5$, and connect $c_4$ to $c_5$, resulting in the graph of Figure~\ref{fig:union_and_inheritance}(f). 
If we apply the inheritance rule first, we remove $c_2$, copy its properties to $c_5$, and connect $c_1$ to $c_5$ through a union relationship. Finally, we apply the union rule and remove $c_1$, connecting $c_4$ to $c_5$ and $c_3$, also resulting in the graph of Figure~\ref{fig:union_and_inheritance}(f). 

{\em (ii) Inheritance and 1:$M$.} We follow a similar strategy to prove that inheritance and 1:$M$ rules are order-independent, enumerating all possible cases in which those two rules may be triggered in the same graph, as shown in Figure~\ref{fig:inheritance_and_1M}(a), (b), (c), (d). This time, as well as in all the remaining cases ($iii$) - ($vi$), the proof is simpler, since there is no alternative intermediate graph involved, if we follow one rule first or another. The only difference is in the set of properties attached to each concept. Again, we assume that the Jaccard similarity between the two concepts connected with an inheritance relationship is less than $\theta_2$, so the inheritance rule is triggered and the properties of the parent concept are copied to the child concept. 

\begin{figure}[htb]
\centering
\includegraphics[width=1\columnwidth]{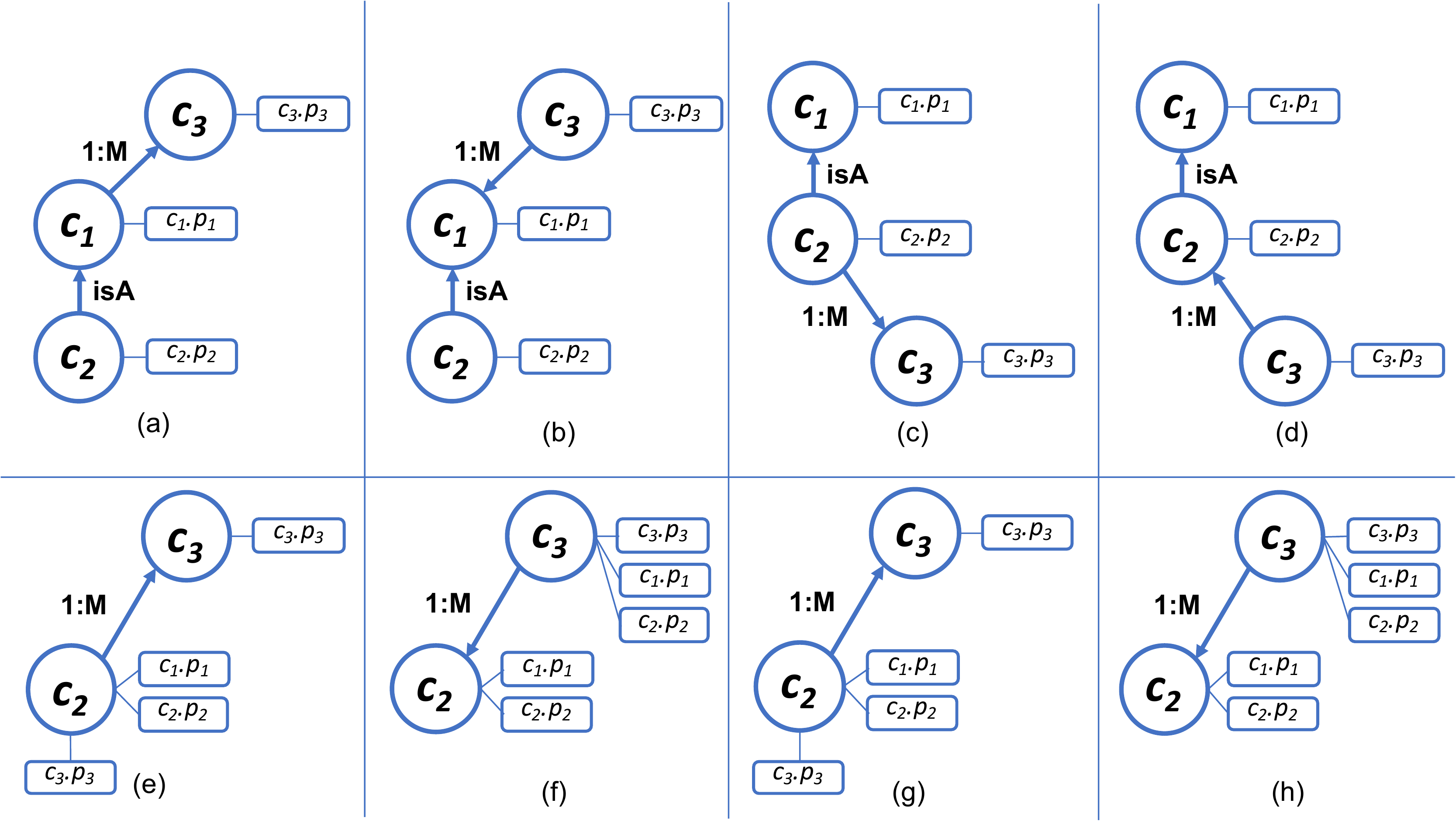}  
\caption{Inheritance and 1:{\em M} Rules Independence.}
\label{fig:inheritance_and_1M}
\end{figure} 

We skip the trivial case in which the inheritance and 1:$M$ relationships are not related, and start with the case depicted in Figure~\ref{fig:inheritance_and_1M}(a), where the parent concept $c_1$ is also the source concept of an 1:$M$ relationship. 
If we apply inheritance first, then we copy the properties of $c_1$ to $c_2$, remove $c_1$ and connect $c_2$ to $c_3$ through a 1:$M$ relationship. Then, we apply the 1:$M$ rule and copy $c_3$'s properties to $c_2$, resulting in the graph of Figure~\ref{fig:inheritance_and_1M}(e).
If we apply the 1:$M$ rule first, then we first copy the properties of $c_3$ to $c_1$ and then we apply inheritance to copy the properties of $c_1$ (also including the properties of $c_3$) to $c_2$, remove $c_1$ and connect $c_2$ to $c_3$ through a 1:$M$ relationship, resulting again in the graph of Figure~\ref{fig:inheritance_and_1M}(e).

In the case of Figure~\ref{fig:inheritance_and_1M}(b), the parent concept ($c_1$) is now also the destination of an 1:$M$ relationship. 
If we apply inheritance first, then we copy the properties of $c_1$ to $c_2$, remove $c_1$ and connect $c_3$ to $c_2$ through a 1:$M$ relationship. Then, we apply the 1:$M$ rule and copy $c_2$'s properties to $c_3$, resulting in the graph of Figure~\ref{fig:inheritance_and_1M}(f).
If we apply the 1:$M$ rule first, then we first copy the properties of $c_1$ to $c_3$ and then we apply inheritance to copy the properties of $c_1$ to $c_2$, remove $c_1$ and connect $c_3$ to $c_2$ through a 1:$M$ relationship. 
Finally, we apply 1:$M$ rule again (remember that Algorithm~\ref{algo:o2p:nospace} iterates until convergence) and and copy the properties of $v_2$ to $v_3$, again resulting in the graph of Figure~\ref{fig:inheritance_and_1M}(f).

In Figure~\ref{fig:inheritance_and_1M}(c), $c_2$ is a child and a source concept of a 1:$M$ relationship. 
In short, if we apply inheritance first, we remove $c_1$ and copy its properties to $c_2$ and then we apply 1:$M$ and also copy the properties of $c_3$ to $c_1$, resulting in Figure~\ref{fig:inheritance_and_1M}(g). 
If we apply 1:$M$ first, we copy the properties of $c_3$ to $c_2$ and then we apply inheritance to copy the properties of $c_1$ to $c_2$ and remove $c_1$, again resulting in Figure~\ref{fig:inheritance_and_1M}(g). 

Finally, in Figure~\ref{fig:inheritance_and_1M}(d), $c_2$ is a child and a destination concept of a 1:$M$ relationship. 
If we apply inheritance first, we remove $c_1$ and copy its properties to $c_2$ and then we apply 1:$M$ and copy the properties of $c_2$ (including the properties of $c_1$) to $c_3$, resulting in the graph of Figure~\ref{fig:inheritance_and_1M}(h). 
If we apply 1:$M$ first, we copy the properties of $c_2$ to $c_3$ and then we apply inheritance to copy the properties of $c_1$ to $c_2$ and remove $c_1$. Again, we need to trigger the 1:$M$ rule once more to copy the properties of $c_2$, now also including the properties of $c_1$, to $c_3$ and get the graph of Figure~\ref{fig:inheritance_and_1M}(h).

For the remaining pairs of rules ($iii$) -- ($vi$), we can follow the same strategy and prove that they are order-independent for all possible cases.

\textbf{Induction hypothesis.} 
Applying the rules in any order for any $R' \subseteq R$, where $|R'|$=$n$, always results in the same $O'$. 

Then, applying the rules in any order for any $R'' \subseteq R$, such that $|R''|$ = $n$+1 and $R' \subset R''$, will always result in the same $O''$, since there is only one additional relationship in $R''$ compared to $R'$, and only one possible rule corresponding to this new relationship can be triggered.
\end{proof}

%% file: appendix2.tex
\section{0/1 Knapsack Problem Reduction}
\label{sec:appendix:1}

\begin{proof}
Given an instance of 0/1 Knapsack problem, our reduction produces the following instance of relationship selection: the cost $Cost(r_i)$ of relationship $r_i$ is set to $w_i$, and the benefit $Benefit(r_i)$ of relationship $r_i$ is set to $b_i$ as well. We set the space limit $S$ to $W$. Clearly this reduction runs in polynomial time. 

If we started with a YES instance of 0/1 Knapsack, then we claim that the reduction produces a YES instance of relationship selection. Suppose there exists a subset $T\subseteq X$ for which $\sum_{i\in T}b_i = B$ is maximized and $\sum_{i\in T}w_i \leq W$. Then selecting the relationship in $T$ has total benefit $B$ and weight no greater than $W$, so the instance of relationship selection produced by the reduction is a YES instance. 

If the reduction produces a YES instance of relationship selection, then we claim that ($X, B$) is a YES instance of 0/1 Knapsack. Let $T\subseteq X$ be the selected relationships, whose total benefit is $B$ and whose total cost is at most $W$. In other words, we have $\sum_{i\in T}b_i = B$ and $\sum_{i\in T}w_i \leq W$. We conclude that ($X, B$) is a YES instance of Knapsack problem as required.  
\end{proof}

%% file: main.bbl
\begin{thebibliography}{10}

\bibitem{neptune}
Amazon neptune.
\newblock \url{https://aws.amazon.com/neptune/}, March 2020.

\bibitem{cosmos}
Azure cosmos db.
\newblock \url{https://azure.microsoft.com/en-us/services/cosmos-db/}, March
  2020.

\bibitem{FDIC}
Federal deposit insurance corporation.
\newblock \url{https://www.fdic.gov/regulations/resources/call/index.html},
  March 2020.

\bibitem{gremlin}
Gremlin query language.
\newblock \url{https://tinkerpop.apache.org/gremlin.html}, March 2020.

\bibitem{janus}
Janusgraph: Distributed graph database.
\newblock \url{http://janusgraph.org/}, March 2020.

\bibitem{neo4j}
The neo4j graph platform.
\newblock \url{https://neo4j.com/}, March 2020.

\bibitem{owl}
Owl 2 web ontology language document overview.
\newblock \url{https://www.w3.org/TR/owl2-overview/}, March 2020.

\bibitem{SEC}
Securities and exchange commission.
\newblock
  \url{https://www.sec.gov/dera/data/financial-statement-data-sets.html}, March
  2020.

\bibitem{DBLP:journals/vldb/AbadiMMH09}
D.~J. Abadi, A.~Marcus, S.~Madden, and K.~Hollenbach.
\newblock Sw-store: a vertically partitioned {DBMS} for semantic web data
  management.
\newblock {\em {VLDB} J.}, 18(2):385--406, 2009.

\bibitem{Agrawal:2000}
S.~Agrawal, S.~Chaudhuri, and V.~R. Narasayya.
\newblock Automated selection of materialized views and indexes in sql
  databases.
\newblock In {\em {VLDB}}, pages 496--505, 2000.

\bibitem{Bollacker08freebase}
K.~D. Bollacker, C.~Evans, P.~Paritosh, T.~Sturge, and J.~Taylor.
\newblock Freebase: a collaboratively created graph database for structuring
  human knowledge.
\newblock In {\em ACM SIGMOD}, pages 1247--1250, 2008.

\bibitem{DBLP:journals/pvldb/BonifatiMT17}
A.~Bonifati, W.~Martens, and T.~Timm.
\newblock An analytical study of large {SPARQL} query logs.
\newblock {\em Proc. {VLDB} Endow.}, 11(2):149--161, 2017.

\bibitem{DBLP:conf/sigmod/BorneaDKSDUB13}
M.~A. Bornea, J.~Dolby, A.~Kementsietsidis, K.~Srinivas, P.~Dantressangle,
  O.~Udrea, and B.~Bhattacharjee.
\newblock Building an efficient {RDF} store over a relational database.
\newblock In {\em {ACM SIGMOD}}, pages 121--132, 2013.

\bibitem{Brin:1998}
S.~Brin and L.~Page.
\newblock The anatomy of a large-scale hypertextual web search engine.
\newblock In {\em {WWW}}, pages 107--117, 1998.

\bibitem{Bruno:2005}
N.~Bruno and S.~Chaudhuri.
\newblock Automatic physical database tuning: A relaxation-based approach.
\newblock In {\em {ACM SIGMOD}}, pages 227--238, 2005.

\bibitem{DBLP:conf/vldb/ChongDES05}
E.~I. Chong, S.~Das, G.~Eadon, and J.~Srinivasan.
\newblock An efficient sql-based {RDF} querying scheme.
\newblock In {\em {VLDB}}, pages 1216--1227, 2005.

\bibitem{DBLP:series/synthesis/2015Christophides}
V.~Christophides, V.~Efthymiou, and K.~Stefanidis.
\newblock {\em Entity Resolution in the Web of Data}.
\newblock Synthesis Lectures on the Semantic Web: Theory and Technology. Morgan
  {\&} Claypool Publishers, 2015.

\bibitem{Codd:1970}
E.~F. Codd.
\newblock A relational model of data for large shared data banks.
\newblock {\em Commun. ACM}, 13(6):377--387, June 1970.

\bibitem{DBLP:conf/icde/TrindadeKCMS20}
J.~M.~F. da~Trindade, K.~Karanasos, C.~Curino, S.~Madden, and J.~Shun.
\newblock Kaskade: Graph views for efficient graph analytics.
\newblock In {\em {ICDE}}, pages 193--204, 2020.

\bibitem{Dash:2011}
D.~Dash, N.~Polyzotis, and A.~Ailamaki.
\newblock Cophy: A scalable, portable, and interactive index advisor for large
  workloads.
\newblock {\em {PVLDB}}, 4(6):362--372, 2011.

\bibitem{DBLP:journals/corr/abs-1901-08248}
A.~Deutsch, Y.~Xu, M.~Wu, and V.~Lee.
\newblock Tigergraph: {A} native {MPP} graph database.
\newblock {\em CoRR}, abs/1901.08248, 2019.

\bibitem{DBLP:series/synthesis/2015Dong}
X.~L. Dong and D.~Srivastava.
\newblock {\em Big Data Integration}.
\newblock Synthesis Lectures on Data Management. Morgan {\&} Claypool
  Publishers, 2015.

\bibitem{DBLP:journals/tkde/FanWW16}
W.~Fan, X.~Wang, and Y.~Wu.
\newblock Answering pattern queries using views.
\newblock {\em {IEEE} Trans. Knowl. Data Eng.}, 28(2):326--341, 2016.

\bibitem{Finkelstein:1988}
S.~Finkelstein, M.~Schkolnick, and P.~Tiberio.
\newblock Physical database design for relational databases.
\newblock {\em ACM Trans. Database Syst.}, 13(1):91--128, 1988.

\bibitem{DBLP:conf/sigmod/FrancisGGLLMPRS18}
N.~Francis, A.~Green, P.~Guagliardo, et~al.
\newblock Cypher: An evolving query language for property graphs.
\newblock In {\em ACM SIGMOD}, pages 1433--1445, 2018.

\bibitem{DBLP:conf/wise/HarrisS05}
S.~Harris and N.~Shadbolt.
\newblock {SPARQL} query processing with conventional relational database
  systems.
\newblock In {\em {WISE}}, pages 235--244, 2005.

\bibitem{Hartig:2019:DSP:3327964.3328495}
O.~Hartig and J.~Hidders.
\newblock Defining schemas for property graphs by using the graphql schema
  definition language.
\newblock In {\em Proceedings of the 2Nd Joint International Workshop on Graph
  Data Management Experiences \& Systems (GRADES) and Network Data Analytics
  (NDA)}, GRADES-NDA'19, pages 6:1--6:11, 2019.

\bibitem{DBLP:conf/edbt/HassanKJAS18}
M.~S. Hassan, T.~Kuznetsova, H.~C. Jeong, W.~G. Aref, and M.~Sadoghi.
\newblock Extending in-memory relational database engines with native graph
  support.
\newblock In {\em {EDBT}}, pages 25--36, 2018.

\bibitem{Huang2011ScalableSQ}
J.~Huang, D.~J. Abadi, and K.~Ren.
\newblock Scalable sparql querying of large rdf graphs.
\newblock {\em PVLDB}, 4:1123--1134, 2011.

\bibitem{Kimura:2010}
H.~Kimura, G.~Huo, A.~Rasin, S.~Madden, and S.~B. Zdonik.
\newblock Coradd: Correlation aware database designer for materialized views
  and indexes.
\newblock {\em PVLDB}, 3(1-2):1103--1113, 2010.

\bibitem{LehmannIJJKMHMK15}
J.~Lehmann, R.~Isele, M.~Jakob, et~al.
\newblock Dbpedia - {A} large-scale, multilingual knowledge base extracted from
  wikipedia.
\newblock {\em Semantic Web}, 2015.

\bibitem{Leskovec:2014}
J.~Leskovec, A.~Rajaraman, and J.~D. Ullman.
\newblock {\em Mining of Massive Datasets}.
\newblock Cambridge University Press, New York, NY, USA, 2nd edition, 2014.

\bibitem{DBLP:conf/www/MadukoASS07}
A.~Maduko, K.~Anyanwu, A.~P. Sheth, and P.~Schliekelman.
\newblock Estimating the cardinality of {RDF} graph patterns.
\newblock In {\em {WWW}}, pages 1233--1234, 2007.

\bibitem{DBLP:conf/icde/MeimarisPMA17}
M.~Meimaris, G.~Papastefanatos, N.~Mamoulis, and I.~Anagnostopoulos.
\newblock Extended characteristic sets: Graph indexing for {SPARQL} query
  optimization.
\newblock In {\em {ICDE}}, pages 497--508, 2017.

\bibitem{7498239}
M.~J. Mior, K.~Salem, A.~Aboulnaga, and R.~Liu.
\newblock Nose: Schema design for nosql applications.
\newblock In {\em {ICDE}}, pages 181--192, May 2016.

\bibitem{DBLP:conf/icde/NeumannM11}
T.~Neumann and G.~Moerkotte.
\newblock Characteristic sets: Accurate cardinality estimation for {RDF}
  queries with multiple joins.
\newblock In {\em {ICDE}}, pages 984--994, 2011.

\bibitem{DBLP:journals/vldb/NeumannW10}
T.~Neumann and G.~Weikum.
\newblock The {RDF-3X} engine for scalable management of {RDF} data.
\newblock {\em {VLDB} J.}, 19(1):91--113, 2010.

\bibitem{Noy:2019}
N.~Noy, Y.~Gao, A.~Jain, A.~Narayanan, A.~Patterson, and J.~Taylor.
\newblock Industry-scale knowledge graphs: Lessons and challenges.
\newblock {\em Commun. ACM}, 62(8):36--43, July 2019.

\bibitem{DBLP:conf/sigmod/QuamarLMOKME20}
A.~Quamar, C.~Lei, D.~Miller, et~al.
\newblock An ontology-based conversation system for knowledge bases.
\newblock In {\em {ACM SIGMOD}}, pages 361--376, 2020.

\bibitem{Abdul:2018}
A.~Quamar, F.~\"{O}zcan, and K.~Xirogiannopoulos.
\newblock Discovery and creation of rich entities for knowledge bases.
\newblock In {\em {ExploreDB}}, 2018.

\bibitem{Robinson:2013}
I.~Robinson, J.~Webber, and E.~Eifrem.
\newblock {\em Graph Databases}.
\newblock O'Reilly Media, Inc., 2013.

\bibitem{DBLP:journals/sigmod/SakrA09}
S.~Sakr and G.~Al{-}Naymat.
\newblock Relational processing of {RDF} queries: a survey.
\newblock {\em {SIGMOD} Record}, 38(4):23--28, 2009.

\bibitem{blind:2018}
J.~Sen, F.~Ozcan, A.~Quamar, G.~Stager, A.~R. Mittal, M.~Jammi, C.~Lei,
  D.~Saha, and K.~Sankaranarayanan.
\newblock Natural language querying of complex business intelligence queries.
\newblock In {\em {SIGMOD}}, pages 1997--2000, 2019.

\bibitem{Suchanek:2008}
F.~M. Suchanek, G.~Kasneci, and G.~Weikum.
\newblock Yago: A large ontology from wikipedia and wordnet.
\newblock {\em Semantic Web}, 2008.

\bibitem{DBLP:conf/sigmod/SunFSKHX15}
W.~Sun, A.~Fokoue, K.~Srinivas, A.~Kementsietsidis, G.~Hu, and G.~T. Xie.
\newblock Sqlgraph: An efficient relational-based property graph store.
\newblock In {\em {ACM SIGMOD}}, pages 1887--1901, 2015.

\bibitem{szarnyas2017incremental}
G.~Sz{\'a}rnyas.
\newblock Incremental view maintenance for property graph queries.
\newblock {\em arXiv preprint arXiv:1712.04108}, 2017.

\bibitem{DBLP:conf/sigmod/TianXZPTSKAP20}
Y.~Tian, E.~L. Xu, W.~Zhao, et~al.
\newblock {IBM} db2 graph: Supporting synergistic and retrofittable graph
  queries inside {IBM} db2.
\newblock In {\em {SIGMOD}}, pages 345--359, 2020.

\bibitem{DBLP:conf/edbt/TsialiamanisSFCB12}
P.~Tsialiamanis, L.~Sidirourgos, I.~Fundulaki, V.~Christophides, and P.~A.
  Boncz.
\newblock Heuristics-based query optimisation for {SPARQL}.
\newblock In {\em {EDBT}}, pages 324--335, 2012.

\bibitem{DBLP:conf/grades/RestHKMC16}
O.~van Rest, S.~Hong, J.~Kim, X.~Meng, and H.~Chafi.
\newblock {PGQL:} a property graph query language.
\newblock In {\em Graph Data-management Experiences and Systems}, page~7, 2016.

\bibitem{Vazirani:2001}
V.~V. Vazirani.
\newblock {\em Approximation Algorithms}.
\newblock Springer-Verlag, Berlin, Heidelberg, 2001.

\bibitem{Zilio:2004}
D.~C. Zilio, J.~Rao, S.~Lightstone, et~al.
\newblock Db2 design advisor: Integrated automatic physical database design.
\newblock In {\em {VLDB}}, pages 1087--1097, 2004.

\end{thebibliography}
